%% file: plasmon_main.tex
\documentclass[aps,prb,showpacs,superscriptaddress,twocolumn,amsmath,amssymb]{revtex4}

\usepackage{graphicx}
\usepackage{dcolumn}
\usepackage{multirow}
\usepackage{color}
\usepackage[normalem]{ulem}

\usepackage{dcolumn}
\usepackage{ifthen}
\usepackage{threeparttable}
\usepackage{multirow}
\usepackage{longtable}
\usepackage{color}
\usepackage[normalem]{ulem}

\include{newcommands}

\begin{document}

\title{{\rm\small\hfill }\\
      First-principles calculations and model analysis of plasmon excitations in graphene}

\author{Pengfei ~Li}
\affiliation{Key Laboratory of Quantum Information, University of Science and Technology of China, Hefei, 230026, China}
\affiliation{Synergetic Innovation Center of Quantum Information and Quantum Physics, University of Science and Technology of China, Hefei, 230026, China}

\author{Xinguo ~Ren}
\affiliation{Key Laboratory of Quantum Information, University of Science and Technology of China, Hefei, 230026, China}
\affiliation{Synergetic Innovation Center of Quantum Information and Quantum Physics, University of Science and Technology of China, Hefei, 230026, China}

\author{Lixin ~He}
\affiliation{Key Laboratory of Quantum Information, University of Science and Technology of China, Hefei, 230026, China}
\affiliation{Synergetic Innovation Center of Quantum Information and Quantum Physics, University of Science and Technology of China, Hefei, 230026, China}

\date{\today }

\begin{abstract}
Plasmon excitations in free-standing graphene and graphene/hexagonal boron nitride (hBN) heterostructure are studied using linear-response time-dependent density functional theory within the random phase approximation. Within a single theoretical framework, we examine both the plasmon dispersion behavior and lifetime (line width) of Dirac and $\pi$ plasmons on an equal footing. Particular attention is paid to the influence of the hBN substrate and the anisotropic effect.
Furthermore, a model-based analysis indicates that the correct dispersion behavior of $\pi$ plasmons should be $\omega_\pi(q) = \sqrt{E_g^2 + \beta q}$ for small
$q$'s, where $E_g$ is the band gap at the $M$ point in the Brillouin zone, and $\beta$ is a fitting parameter. This model is radically different from previous
proposals, but in good agreement with our calculated results from first principles.
\end{abstract}

\maketitle

\section{\label{sec:intro}Introduction}


In recent years the plasmonic excitations in graphene and graphene-related materials have attracted considerable attention both theoretically \cite{Huang/Lin/Chuu:1997,Shyu/Lin:2000,Wunsch/etal:2006,Hwang/DasSarma:2007,Jablan/etal:2009,Yan/etal:2011} and experimentally \cite{Eberlein/etal:2008,Kramberger/etal:2008,Liu/etal:2008,Lu/etal:2009,Koppens/etal:2011,Kinyanjui/etal:2012,Grigorenko/etal:2012,Woessner/etal:2015,Liou/etal:2015} due to their importance for basic physics and technological applications.
 Derived from the unique electronic band structure, the plasmon excitation spectra of graphene span a wide energy range, and fall into three distinct regimes. At low energies (0-2 eV) with finite electron doping, graphene can sustain the so-called Dirac plasmons, originating from the intraband transitions of Dirac fermions in the vicinity of $K$ points of the Brillouin zone (BZ) \cite{Wunsch/etal:2006,Hwang/DasSarma:2007}. At higher energies (4-15 eV), there exist the intrinsic $\pi$ plasmons in graphene due to the collective excitations of electrons from $\pi$ to $\pi^\ast$ bands\cite{Marinopoulos/etal:2004}.
The $\pi$ plasmons are very dispersive, starting at $\sim 4$ eV for small momentum transfer $q$ all the way to $\sim 15$ eV for the approaching to the boundary of the BZ.  At even higher energies, $\sigma$ bands start to contribute, the mixture of the $\pi \rightarrow \pi^\ast$  and $\sigma$ transitions leads to yet another set of distinct
plasmon peaks, usually denoted as $\pi$+$\sigma$ plasmons. As such, the rich plasmon physics in graphene renders the system distinguished from normal metals, conventional 2-dimensional electron gas, and doped semiconductors.

The low-energy Dirac plasmon excitations are only present under finite electron or hole dopings \cite{Wunsch/etal:2006,Hwang/DasSarma:2007}, which can be readily achieved by chemical means or by electric gating.This type of plasmons attract enormous interest for potential technological applications due to
the consideration given to both strong field localization and low energy loss, simultaneously\cite{Jablan/etal:2009,Low/Avouris:2014}. Furthermore, they are highly flexible in the sense that their frequencies can be tuned from Terahertz to middle infrared by varying the doping level, and can be ``engineered" by encapsulating graphene in between other 2D layered materials \cite{Woessner/etal:2015}. Because of this unique property, graphene has been considered as a promising material for fabricating nano-plasmonic devices.  A review on the recent progress and future perspective of this field can be found in Ref.~[\onlinecite{Low/Avouris:2014}]. The $\pi$ and $\pi$+$\sigma$ plasmons at higher energies are also of significant scientific interest and  have been under intensive theoretical \cite{Huang/Lin/Chuu:1997,Yan/etal:2011,Kramberger/etal:2008} and experimental \cite{Lu/etal:2009,Koppens/etal:2011} investigations. It should be noted that these latter types of plasmons are also present in the parent material 
of graphene -- graphite.
However,it shows that the dispersion behavior in graphene is much different from the graphite's, especially in small $q$. Consequently, the interest here is often on the dispersion behavior of the plasmon peaks \cite{Kramberger/etal:2008,Lu/etal:2009,Liou/etal:2015} as a function of the momentum transfer $q$, as well as the influence of the substrates \cite{Yan/etal:2011} and/or the interlayer interactions within graphene-based heterostructures or multi-layer graphenes.
\cite{Koppens/etal:2011}.


Historically, researches on the above-mentioned different plasmon types have been largely conducted
within different communities. Theoretically, the studies of Dirac plasmons have been predominantly carried out with the linear band model (also known as
``Dirac-cone approximation") \cite{Wunsch/etal:2006,Hwang/DasSarma:2007,Jablan/etal:2009},  which is valid only in the vicinity of the $K$ points in the BZ.
Nevertheless, recent \textit{ab initio} studies on Dirac plasmons based on time-dependent density-functional theory (TDDFT) revealed the appreciable non-isotropic
effect (the variation of the plasmon dispersions along different directions in the Brillouin zone) \cite{Gao/Yuan:2011,Pisarra/etal:2014} and the existence
of acoustic Dirac plasmon mode in graphene \cite{Pisarra/etal:2014}. Physics of this type such cannot be captured
by the linear band model.  The studies of $\pi$ plasmons, on the other hand, have been mainly done with \textit{ab-initio} TDDFT within the
random-phase approximation (RPA) \cite{Kramberger/etal:2008,Yan/etal:2011,Liou/etal:2015}, except for some earlier ones based on tight-binding model involving
only $\pi$ bands \cite{Huang/Lin/Chuu:1997,Shyu/Lin:2000}. Compared to other two types of plasmons, $\pi$+$\sigma$ plasmons have received less attention,
and the study of these excitations would require including high-lying energy bands. Hence for $\pi$+$\sigma$ plasmons, the \textit{ab-initio} approach would be
more appropriate. All together, it
appears that TDDFT-RPA can offer a unified description of all types of plasmon excitations in graphene and graphene-related materials in a fully \textit{ab initio}
manner, thus eliminating possible ambiguities arising from the choice of model parameters and/or approximations that might miss important physical effects.

Despite the intensive studies of the plasmon excitations in graphene, some issues remain unclear and await for further investigations. For instance,
the dispersion behavior of $\pi$ plasmons has long been thought to be linear \cite{Kramberger/etal:2008,Yan/etal:2011,Lu/etal:2009,Kinyanjui/etal:2012}, but
this view was challenged by Liou \textit{et al.} \cite{Liou/etal:2015} recently. These authors claimed that $\pi$ plasmons follow a $\sqrt{q}$ behavior for small
$q$'s, similar to the case of Dirac plasmons.
In this work, we present a comprehensive study of Dirac and $\pi$ plasmon excitations in graphene and graphene/monolayer hexagonal boron nitride (hBN) heterostructure
 using \textit{ab initio} TDDFT-RPA approach. We examine both the dispersion behavior (plasmon peak positions) and the line-width (plasmon lifetimes) as a
function of the momentum transfer $\bfq$. Our \textit{ab initio} calculations are further complemented by model analysis, which allows us to gain new insight into 
the dispersion behavior of $\pi$ plasmons. For the lifetimes, \textit{ab initio} results from TDDFT-RPA have not been reported in the literature.
In this connection, we note that only the Landau damping channel is adequately treated within the standard RPA. Effects from impurities, disorders, and
electron-phonon couplings are usually interpreted via the framework of RPA combined with the number conserving relaxation time (RT)
approximation\cite{Mermin:1970,Jablan/etal:2009}. In this approach, the relaxation time due to the scattering from impurities or phonons is estimated empirically and
put into the RPA-RT framework by hand. It appears a full \textit{ab initio} treatment of the plasmon lifetime problem taking into all important physical channels
is still not practical. In this work, our \textit{ab initio} lifetime calculations are restricted to the level of electron-electron scattering within RPA.
A full account of other contributions from first principles will be pursued in future work.

The rest of the paper is organized as follows. In Sec.~\ref{sec:methods} the basic equations of the TDDFT-RPA approach behind our implementation are presented.
This is followed by a description of the implementation details and computational setups in Sec.~\ref{sec:implementation}. The calculated results for graphene and
graphene/hBN are then presented in Sec.~\ref{sec:results}, complemented by an in-depth model analysis and discussions. Finally we summarize our work in Sec.~\ref{sec:conclusion}.

\section{\label{sec:methods}Methods}
	
The plasmon excitations can be measured experimentally by the momentum-resolved electron energy loss spectroscopy (EELS) technique. Theoretically and computationally, TDDFT-RPA represents a powerful \textit{ab initio} approach to describe EELS, as thoroughly discussed in Ref.~[\onlinecite{Onida/Reining/Rubio:2002,Silkin/Chulkov/Echenique:2004,Yuan/Gao:2009,Mowbray:2014}]. A most recent analysis of the role of theoretical dielectric function models in describing EELS can be found in Ref.~[\onlinecite{Azzolini/etal:2017}]. Here we present the key equations of the formalism as a basis for the pertinent technical details behind our implementations.   In brief, EELS can be obtained from the inverse of the dielectric function $\varepsilon$, or equivalently the linear charge density response function $\chi$,
\begin{equation}
-\text{Im}\{\varepsilon^{-1}(\textbf{q},\omega)\} =  -{4\pi \over {q^2}}\text{Im}\{\chi_{\textbf{G}=0,\textbf{G}^\prime=0}(\textbf{q},\omega)\} \, .
 \label{eq:loss_function}
\end{equation}
%
where $\bfG$ and $\bfGp$ are the 3-dimensional (3D) reciprocal lattice vectors, $\bfq$ is a wave vector within the first BZ (1BZ), and $\omega$ is the frequency.
Here only the ``head" term ($\bfG=\bfGp=0$) of $\varepsilon^{-1}$ is of our concern, which corresponds to the response function of the system at the macroscopic scale.
Within TDDFT, the system's interacting response function $\chi$ is linked to its non-interacting counterpart $\chi_0$ via the Dyson
equation:
\begin{widetext}
\begin{equation}
\chi_{\textbf{G},\textbf{G}^\prime}(\textbf{q},\omega) = \chi_{\textbf{G},\textbf{G}^\prime}^0(\textbf{q},\omega) + \sum_{\textbf{G}_1,\textbf{G}_2}\chi^0_{\textbf{G},\textbf{G}_1}(\textbf{q},\omega)K_{\textbf{G}_1,\textbf{G}_2}(\textbf{q},\omega)\chi_{\textbf{G}_2,\textbf{G}^\prime}(\textbf{q},\omega),
\label{eq:LR-TDDFT}
\end{equation}
\end{widetext}
where the non-interacting response function $\chi_0$ is given explicitly by the well-known Adler-Wiser formula \cite{Adler:1962,Wiser:1963}
\begin{widetext}
\begin{equation}
\chi^0_{\textbf{G},\textbf{G}^\prime}(\textbf{q},\omega) = {1 \over \Omega} \sum^\text{1BZ}_\textbf{k} \sum_{n,n^\prime}{{f_{n,\textbf{k}}-f_{n^\prime,\textbf{k}+\textbf{q}}} \over {\omega + \epsilon_{n,\textbf{k}} -\epsilon_{n^\prime,\textbf{k}+\textbf{q}} + i\eta}} \langle n,\textbf{k}|e^{-i(\textbf{q}+\textbf{G})\textbf{r}}|n^\prime,\textbf{k}+\textbf{q}\rangle \langle n^\prime,\textbf{k}+\textbf{q}|e^{i(\textbf{q}+\textbf{G}^\prime)\textbf{r}^\prime}|n,\textbf{k}\rangle,
\label{eq:chi_0}
\end{equation}
\end{widetext}
In Eq.~(\ref{eq:chi_0}), $\Omega$ stands for the volume of the Born-von-Karmen supercell, $f_{n,k}$, $\epsilon_{n,k}$, $|n,\bfk \rangle$
are the Fermi occupation numbers, Kohn-Sham (KS) \cite{Kohn/Sham:1965} eigenvalues, and KS eigenvectors, respectively. The computation
of $\chi^0_{\textbf{G},\textbf{G}^\prime}(\textbf{q},\omega)$ will be discussed in the next section.
Within RPA, the full kernel $K_{\textbf{G}_1,\textbf{G}_2}(\textbf{q},\omega)$ in Eq~(\ref{eq:LR-TDDFT}) is reduced to the static
Coulomb kernel,
\begin{equation}
K^C_{\textbf{G}_1,\textbf{G}_2}(\textbf{q}) = {4\pi \over |\textbf{q} + \textbf{G}_1|^2} \delta_{\textbf{G}_1,\textbf{G}_2}\, .
\label{Eq:2D_Coulomb_kernel}
\end{equation}
The above formalism (Eqs.~\ref{eq:LR-TDDFT}-\ref{Eq:2D_Coulomb_kernel}) is perfectly suitable for 3D periodic bulk materials, but needs modifications for 2-dimensional (2D) materials as explained below. In the 2D case, the system is infinite and periodic only in the basal ($x$-$y$) plane, but confined in the third ($z$) direction. 
A so-called supercell approach is commonly
used to treat 2D systems, in which the system is modeled by repeated 2D slabs, separated by a large vacuum region in the $z$ direction.
The advantage of the supercell approach is that the 3D formalism presented above can still be used, but care must be taken to remove spurious interactions
between the periodic replicas. The issue is already well-known in semi-local DFT calculations for polar surfaces. The situation is more severe in
the present case, because of the explicit presence of long-range Coulomb interaction in Eq.~(\ref{eq:LR-TDDFT}). 

In the literature, two schemes have been employed to deal with the above-mentioned problem.  The first scheme,
introduced by Rozzi \textit{et al.} \cite{Rozzi/etal:2006}, sticks to the reciprocal-space formalism, but employs the Fourier transform
of a truncated Coulomb potential, instead of the bare one (Eq. \ref{Eq:2D_Coulomb_kernel}).  Specifically, by restricting the Coulomb interaction within 
a window $[-R, R]$ in the $z$ direction,
the Coulomb kernel in Eq.~(\ref{Eq:2D_Coulomb_kernel}) becomes
\begin{widetext}
\begin{equation}
\tilde{K}^C_{\textbf{G}_1,\textbf{G}_2}(\textbf{q}) = \frac{4\pi \delta_{\bf{G}_1,\bf{G}_2}}{|\bfq+{\bfG}_1|^2}
                         \left[1 + e^{-|\bfq+\bar{\bfG}_1|R}\left(\frac{G_{1,z}}{|\bfq+\bar{\bfG}_1|}\text{sin}(G_{1,z} R) - \text{cos}(G_{1,z}R) \right) \right]
\label{Eq:truncated_Comlomb kernel}
\end{equation}
\end{widetext}
where $\bfG_1=(\bar{\bfG}_1,G_{1,z})$ and $\bfq=(\bar{\bfq},0)$, with $\bar{\bfG}_1$ and $\bar{\bfq}$ being respectively the two-dimensional reciprocal lattice vector
and Bloch wavevector in the basal plane. It has been suggested in Ref.~[\onlinecite{Rozzi/etal:2006}] to choose $R=L_z/2$ where $L_z$ is the length of the lattice
vector in the $z$ direction, and hence $G_z=2\pi n_z/ L_z$ with $n_z$ being an integer number. With this choice, the truncated Coulomb kernel simplifies to,
%
\begin{equation}
\tilde{K}^C_{\textbf{G}_1,\textbf{G}_2}(\bar{\bfq}) = \frac{4\pi\delta_{\textbf{G}_1,\textbf{G}_2}}{|\bar{\bfq}+\textbf{G}_1|^2}
  \left[1 - (-1)^{n_z}e^{-|\bar{\bfq}+\bar{\textbf{G}}_{1}|{L_z \over 2}}\right]
\label{Eq:truncated_Comlomb kernel_simple}
\end{equation}

Now, by replacing the full Coulomb kernel by the truncated one $\tilde{K}^C_{\textbf{G}_1,\textbf{G}_2}(\bar{\bfq})$ in
Eqs.~(\ref{eq:loss_function}),  one obtains the modified 3D interacting response function $\tilde{\chi}$. Finally the loss function of a genuine 2D material
is computed as,
 \begin{equation}
   -\text{Im}\{\epsilon^{-1}_\text{2D}(\bar{\bfq},\omega)\} =  -{4\pi(1-e^{-\bar{q}L_z/2}) \over {{\bar{q}}^2}}\text{Im}\{\tilde{\chi}_{\textbf{G}=0,\textbf{G}^\prime=0}(\bar{\bfq},\omega)\} \, ,
  \label{eq:Imchi-1}
 \end{equation}
where $\bar{q}=|\bar{\bfq}|$. 
It is easy to see that the Fourier transform of the truncated Coulomb interaction approaches its 2D form $2\pi/\bar{q}$ for $\bar{q} << 1/L_z$.
For plasmon excitations in graphene, it has been shown recently by Mowbray \cite{Mowbray:2014} that the artificial interactions between periodic replicas
can affect the plasmon peak positions and intensities quite significantly. Such effects can be efficiently removed by employing the truncated Coulomb
kernel (\ref{Eq:truncated_Comlomb kernel_simple}).
Please note that, for consistency, in Eq.~(\ref{eq:Imchi-1}) we also used the truncated Coulomb interaction form $4\pi(1-e^{-\bar{q}L_z/2}/\bar{q}^2$ 
for the prefactor.  Different from our choice, however, in literature the full Coulomb potential $\frac{4\pi}{{\bar{q}}^2}$ is often used for 
the prefactor in Eq.~(\ref{eq:Imchi-1}).
Our numerical tests show that, using the truncated Coulomb potential here instead of the full one for the prefactor will not produce noticeable difference in the loss spectrum except for extremely small $\bar{q}$'s. Even in the small $\bar{q}$ regime, only the height of the plasmon peaks is slightly affected, but 
not the peak positions.
%

An alternative approach to deal with the artificial interaction issue, as employed by Silkin and coauthors
\cite{Silkin/Chulkov/Echenique:2004,Bergara/etal:2003} and also
by Yuan and Gao \cite{Yuan/Gao:2009,Gao/Yuan:2011}, is a mixed representation of the TDDFT-RPA equations.
In this approach, a reciprocal-space representation for the in-plane ($x$-$y$) dimensions, and a real-space representation for
the out-of-plane ($z$) dimension are employed.
In such a mixed representation, the Dyson equation for the response function (Eq.~\ref{eq:LR-TDDFT}) becomes,
  \begin{widetext}
    \begin{equation}
       \chi_{\bar{\bfG},\bar{\bfG}'}(z,z',\bar{\bfq},\omega) = \chi_{\bar{\bfG},\bar{\bfG}'}^0(z,z',\bar{\bfq},\omega) + \sum_{\bar{\bfG}_1,\bar{\bfG}_2}
          \iint dz_1 dz_2 \chi^0_{\bar{\bfG},\bar{\bfG}_1}(z,z_1,\bar{\bfq},\omega)K_{\bar{\bfG}_1,\bar{\bfG}_2}(z_1,z_2,\bar{\bfq})
               \chi_{\bar{\bfG}_2,\bar{\bfG}'}(z_2,z',\bar{\bfq},\omega),
     \label{eq:LR-TDDFT_mixed}
    \end{equation}
  \end{widetext}
where the Coulomb kernel is given by
  \begin{equation}
      K^C_{\bar{\bfG}_1,\bar{\bfG}_2}(z_1,z_2,\bar{\bfq})=\frac{2\pi \delta_{\bar{\bfG}_1,\bar{\bfG}_2}}{|\bar{\bfq}+\bar{\bfG}_1|}
      e^{-|\bar{\bfq}+\bar{\bfG}_1||z_1-z_2|} \, .
  \end{equation}
In this way, one is essentially dealing with an isolated 2D system, and
the non-physical interactions between periodic images in the 3D reciprocal space formalism are naturally absent.
In practice, however, the mixed representation of the non-interacting response function is often obtained from its 3D reciprocal-space form,
  \begin{equation}
     \chi_{\bar{\bfG},\bar{\bfG}'}^0(z,z',\bar{\bfq},\omega) = \sum_{G_z,G'_z}e^{-iG_z z} \chi_{\bfG,\bfGp}^0(\bar{\bfq}) e^{iG'_z z'}\, .
  \end{equation}
Consequently, $\chi^0$ becomes \textit{de facto} periodic in the $z$ direction of periodicity length $L_z$. Because of this feature, the integration over 
$z$ and $z'$ in Eq.~\ref{eq:LR-TDDFT_mixed} has to be restricted within $(-L_z/2, L_z/2)$. The obtained $\chi_{\bar{\bfG},\bar{\bfG}'}(z,z')$
can then be Fourier transformed to its 3D reciprocal-space form $\chi_{\bfG,\bfGp}$, and the loss spectrum is again computed using
Eq.~(\ref{eq:Imchi-1}). As such, the above two schemes to remove artificial interactions between periodic images become essentially
equivalent \cite{Pisarra/etal:2016}. In the present work, we directly follow the first scheme
(Eqs.~\ref{eq:LR-TDDFT},\ref{eq:chi_0},\ref{Eq:truncated_Comlomb kernel_simple},\ref{eq:Imchi-1}) in our implementation.

\section{\label{sec:implementation}Implementation Details}

One efficient way to compute $\chi^0$ is to first calculate its imaginary part $\chi^S_{\textbf{G},\textbf{G}^\prime}(\textbf{q},\omega)$,
\begin{widetext}
\begin{equation}
\chi^S_{\textbf{G},\textbf{G}^\prime}(\textbf{q},\omega) = {1 \over \Omega} \sum^\text{1BZ}_\textbf{k} \sum_{n,n^\prime}
(f_{n,\textbf{k}}-f_{n^\prime,\textbf{k}+\textbf{q}})\delta(\omega + \epsilon_{n,\textbf{k}} - \epsilon_{n^\prime,\textbf{k}+\textbf{q}})
 \langle n,\textbf{k}|e^{-i(\textbf{q}+\textbf{G})\textbf{r}}|n^\prime,\textbf{k}+\textbf{q} \rangle \langle n^\prime,\textbf{k}+\textbf{q}|e^{i(\textbf{q}+\textbf{G}^\prime)\textbf{r}^\prime}|n,\textbf{k} \rangle,
\label{Eq:chi_0_spectrum}
\end{equation}
\end{widetext}
and then obtain the full $\chi^0$ via the Hilbert transform,
\begin{widetext}
\begin{equation}
\chi^0_{\textbf{G},\textbf{G}^\prime}(\textbf{q},\omega) = \int_0^\infty d\omega^\prime[{1 \over{\omega - \omega^\prime + i\eta}} - {1 \over{\omega + \omega^\prime + i\eta}}]\chi^S_{\textbf{G},\textbf{G}^\prime}(\textbf{q},\omega^\prime).
\label{Eq:chi_0_KK}
\end{equation}
\end{widetext}
Following Ref.~[\onlinecite{Shishkin/Kresse:2006}], the $\delta$-function in Eq.~(\ref{Eq:chi_0_spectrum}) is approximated by
a triangular function,
 \begin{equation}
   \delta(\omega_i-\Delta) \approx  \left\{\begin{array}{ll}
                                           \frac{\Delta - \omega_{i-1}}{\omega_i-\omega_{i-1}}, & \text{for}~~ \omega_{i-1} < \Delta <\omega_i \\
                                                &  \\
                                           \frac{\omega_{i+1} - \Delta}{\omega_{i+1}-\omega_{i}}, & \text{for}~~ \omega_{i} < \Delta <\omega_{i+1} \\
                                         \end{array}
                                   \right.
 \end{equation}
where $\Delta=\epsilon_{n^\prime,\textbf{k}+\textbf{q}}-\epsilon_{n,\textbf{k}}$, and $\omega_i$ is the frequency grid point.
Furthermore, To obtain converged values for the peak positions and width of the loss spectrum, technical parameters
such as the $\bfk$-point mesh, the frequency grid points, and the actual values for the positive infinitesimal parameter $\eta$
must be chosen carefully.

The linear-response TDDFT-RPA equations presented above have been implemented in a recently released first-principles code package
\textit{Atomic-orbital Based Ab-initio Computations at UStc} (ABACUS) \cite{Chen/Guo/He:2010,Li/Liu/etal:2016,abacusweb}.
The Troullier-Martins \cite{Troullier/Martins:1991} norm-conserving
pseudopotential in its fully separable form \cite{Kleinman/Bylander:1982} is used to describe the interactions between nuclear ions and valence electrons.
A ``dual basis set" strategy is adopted in ABACUS which allows to use both plane waves and numerical atomic orbitals (NAOs) bases. In this work,
the plane-wave basis sets are employed to expand the valence-electron wave functions in Eqs.~(\ref{eq:chi_0}) and (\ref{Eq:chi_0_spectrum}).
The Perdew-Zunger local-density approximation (LDA) \cite{Perdew/Zunger:1981} is used for the preceding Kohn-Sham (KS)-DFT \cite{Kohn/Sham:1965}
calculations to obtain the KS orbitals and orbital energies. Unless otherwise stated, the production calculations in this work
are done with the plane-wave basis with a cutoff energy of 50 Rydberg (Ry).

As mentioned in Sec.~\ref{sec:methods}, the supercell approach is used here to model the 2D systems. The distance ($L_z$ in
Eq.~\ref{Eq:truncated_Comlomb kernel}) between periodic layers (slabs) is chosen to be 20 \AA. This choice, together
with the truncated Coulomb potential technique discussed in  Sec.~\ref{sec:methods}, ensures a clean removal of the
artificial interactions between periodic images in the $z$ direction.
In this work, we study two types of systems: (a) a single freestanding graphene and a hBN layer, and (b) a combined graphene/hBN double-layer
heterostructure in AA stacking. For the former, 20 bands (4 occupied plus 16 unoccupied) are included in the calculations of $\chi^S$ in Eq.~(\ref{Eq:chi_0_spectrum}); for the latter, 30 bands 
(8 occupied plus 22 unoccupied) are used. These choices guarantees a convergence of the plasmon spectra up to 30 eV, thus covering the entire energy range of $\pi+\sigma$ plasmons.

In the plasmon dispersion calculations, the positive infinitesimal parameter $\eta$ in Eq.~(\ref{Eq:chi_0_spectrum}) is set to be 0.01Ry, and the $\chi$ matrix is expanded in
terms of 100 $\bfG$-vectors. 
The Brillouin zone is sampled with a 192$\times$192$\times$1 Monkhorst-Pack $\bfk$-point grid. Furthermore, we use a uniform frequency grid with energy spacing of 
0.001 Ry up to 4 Ry, containing 4000 grid points. Extensive tests show that such a combination of parameters are sufficient to obtain converged 
results of the dispersion behavior of
the plasmon excitations.  For lifetimes, highly accurate numbers can only be obtained by extrapolating $\eta \rightarrow 0$.  Benchmark tests on the convergence 
behavior of lifetime calculations will be shown in Appendix~\ref{app:FWHM}. 

In this work, we study the plasmon behavior of extrinsic graphene systems with finite electron doping. To this end, the Fermi level is shifted upwards by 
0.05Ry(0.68eV) above the
Dirac point, and this corresponds to a doping level of 0.027 electrons per unit cell (or free charge carrier concentration of $5.1 \times 10^{13}$ cm$^{-2}$) and the 
resultant Fermi vector is $k_F$ = 0.127 \AA$^{-1}$ as determined from the LDA band structure. In doing so, the tiny effect of doping on the graphene electronic
structure is neglected. Such a procedure is also followed by Pisarra \textit{et al.} in Refs.~[\onlinecite{Pisarra/etal:2014,Pisarra/etal:2016}].
Please note that the Fermi velocity is slightly underestimated within LDA, and correspondingly the the Fermi vector is slightly overestimated for a given doping.
But this does not affect the discussions in the present work.

\section{\label{sec:results}Results}
In this section, the computed results of the full plasmon spectra of graphene and graphene/hBN will be presented.
We will discuss both the dispersion relations and lifetimes. The influence of the hBN substrate
on plasmons in graphene will be also examined.
\subsection{The plasmon dispersion behavior}
\subsubsection{Overall features}
\begin{figure}[h]
\centering
\includegraphics[width=0.45\textwidth]{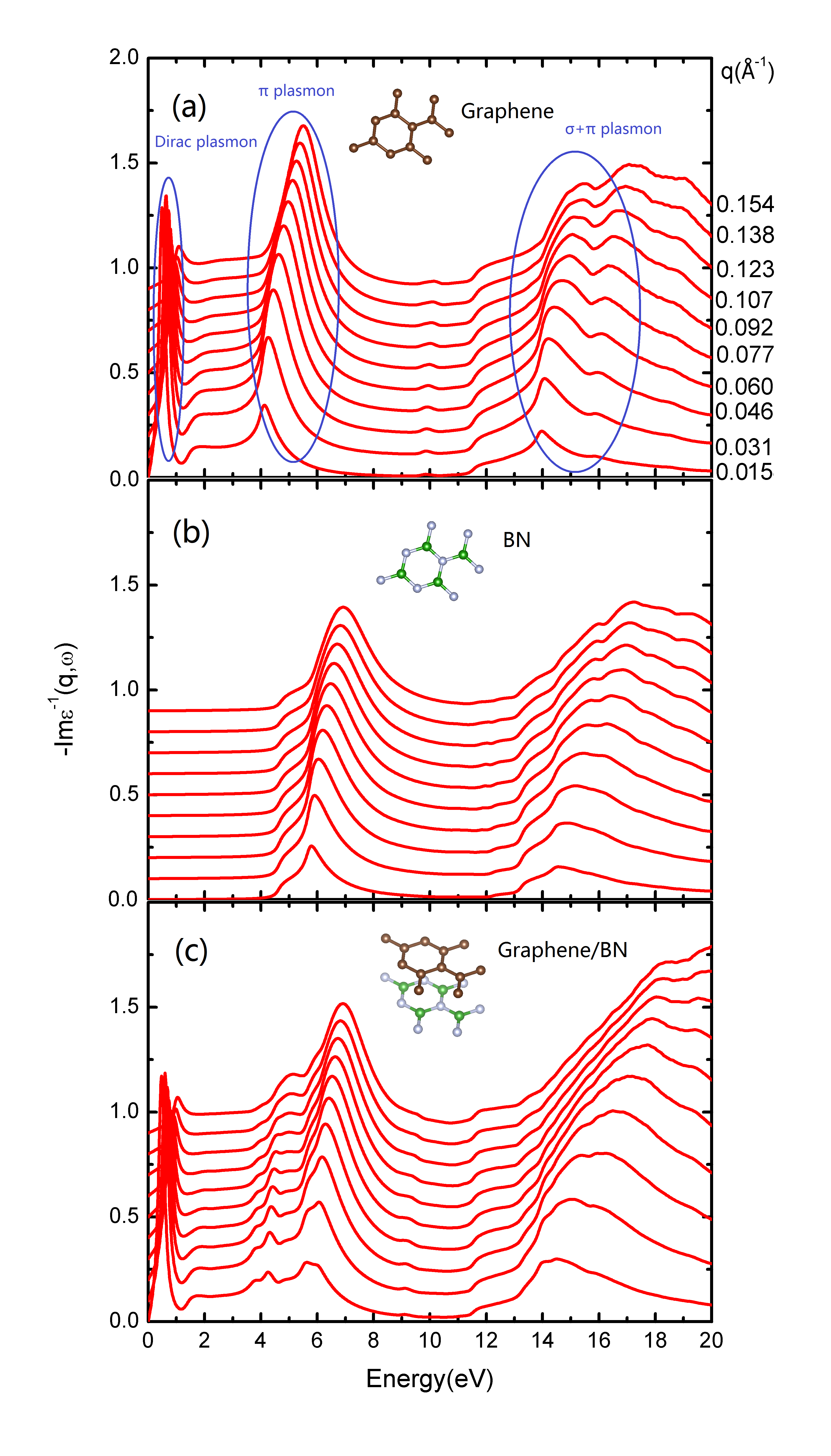}
\caption {The loss spectra for graphene (a), hBN (b), and graphene/hBN bilayer (c) systems for different momentum transfer $q$
along the $\Gamma-M$ direction. The Fermi level is shifted up by 0.05 Ry in graphene and graphene/hBN calculations to mimic the effect of
finite dopings. The three different (Dirac, $\pi$, and $\pi$+$\sigma$) plasmon modes in graphene are labeled in panel (a).} 
\label{fig:whole_spectra}
\end{figure}

In Fig~\ref{fig:whole_spectra} the loss spectra computed using the linear-response TDDFT-RPA approach are presented both for free-standing graphene (upper panel,
Fig.~\ref{fig:whole_spectra}(a)) and the mixed graphene/hBN bilayer (in AA stacking) system (lower panel, Fig.~\ref{fig:whole_spectra}(c)). The result for 
a free-standing hBN layer is also shown (middle panel, Fig.~\ref{fig:whole_spectra}(b)) for comparison. The
different spectral lines aligned vertically correspond to different momentum transfer $\bfq$ along the
$\Gamma-M$ direction in the 1BZ. For calculations of graphene and graphene/hBN, the Fermi level is shifted upwards by 0.05 Ry above the Dirac point to mimic the
effect of finite electron doping. The results for graphene/hBN presented in this section are for AA stacking. However, as shown in
Appendix~\ref{app:g-hBN_stacking_comp}, other types of stacking yield essentially the same results.

The obtained loss functions of doped graphene and graphene/hBN clearly show three distinct plasmon modes, corresponding respectively to
the Dirac plasmons, $\pi$ plasmons, and $\pi$+$\sigma$ plasmons from low to high excitation energies.
The Dirac plasmon peaks are very sharp and pronounced, while $\pi$ plasmon peaks are much broader and carry more spectral weights. The peaks of $\pi$+$\sigma$ plasmons
are even broader, and multiple sub-peak structures within this regime are clearly visible.  As mentioned in the Introduction,
the low-energy Dirac plasmon excitations are only present for extrinsic graphene with finite doping, but the $\pi$ and $\pi$+$\sigma$
are intrinsic and can be activated at both finite and zero dopings.

The comparison between Fig.~\ref{fig:whole_spectra}(a) and (c) clearly reveals the major effect of a single-layer hBN substrate on the
plasmon dispersion behavior of graphene. While the Dirac plasmons are not much affected, $\pi$ plasmons are significantly changed. An overall binodal
peak structure, running parallel to each other as a function of $\bfq$, can be recognized. Comparing the graphene/hBN spectra (Fig.~\ref{fig:whole_spectra}(c)) to the individual graphene (Fig.~\ref{fig:whole_spectra}(a)) and hBN (Fig.~\ref{fig:whole_spectra}(b))
ones, further suggests that the low-energy peaks stem from graphene, though substantially suppressed, while the high-energy peaks originate from hBN, staying
essentially the same as their counterparts in the isolated hBN layer.  In this context, we note that hBN, a wide-gap insulator, sustains plasmon excitations
which behave similarly to the $\pi$ plasmons in graphene, though blue-shifted to higher energies, as can be seen from Fig.~\ref{fig:whole_spectra}(b).
The suppression of $\pi$ plasmons of graphene upon including a substrate was also observed in graphene/SiC, as reported in Ref.~[\onlinecite{Yan/etal:2011}].

\begin{figure}[h]
\centering
\includegraphics[width=0.45\textwidth]{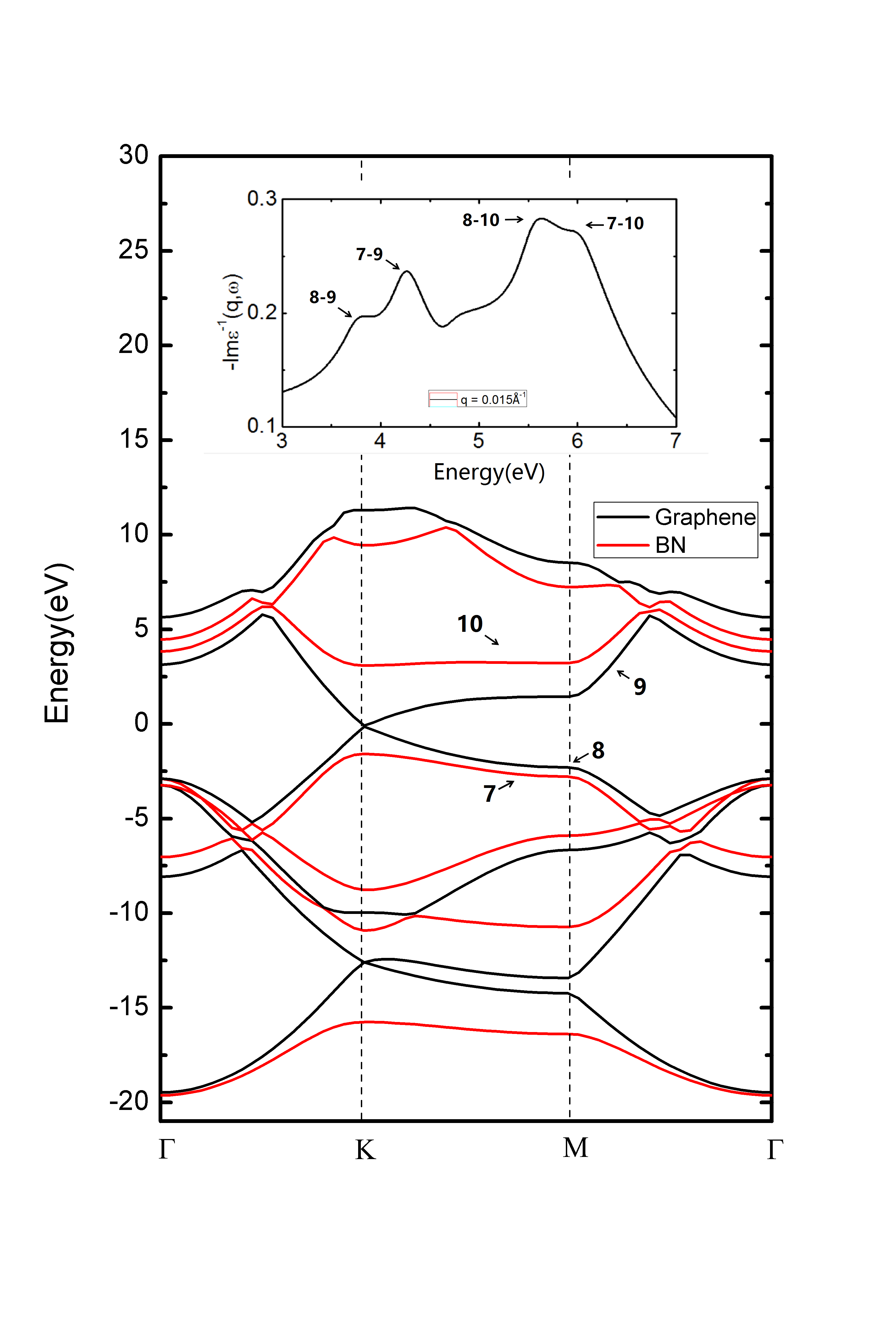}
\caption {(Color online.) Electronic band structure of graphene/hBN in AA stacking, computed with DFT-LDA. Bands in black stems from graphene, while
those in red from hBN.  Inset: a zoom-in of the loss spectra of the hybrid system at a very small momentum value $q=0.015$\AA$^{-1}$. The 4 four
plasmon sub-peaks can be clearly associated with the inter-band transitions around the $M$ point involving energy bands close to the Fermi level, as labeled
in the figure.}
\label{fig:ghBN_band_structure}
\end{figure}

One may notice that, at small $\bfq$'s, there are finer plasmon peak structures within the $\pi$-region of the graphene/hBN heterostructure.
To unravel the origin of these sub-peaks, we plot in Fig.~\ref{fig:ghBN_band_structure} the band structure of graphene/hBN,
obtained with DFT-LDA. Also shown is the zoom-in of the plasmon peaks at $q=0.015$\AA$^{-1}$ within an energy window of $3$ - $7$ eV.
It can be seen that the band structure of hBN is actually very similar to graphene around the $M$ point in BZ, with the direct gap (at $M$) of hBN
about 2 eV larger than that of graphene.  A close inspection of the two graphs further reveals that each
small individual peak can be associated with an inter-band transition between the two occupied bands and two unoccupied bands of the
hybrid graphene/hBN system around the $M$ point, as labeled in Fig.~\ref{fig:ghBN_band_structure}. For example, the first peak
(cf. the inset of Fig.~\ref{fig:ghBN_band_structure}) with an energy of 3.71 eV corresponds well to the transition from the 8-th to 9-th band, while the
third peak of 5.50 eV corresponds to the $8 \rightarrow 10$ interband transition, and so on.

The similarity between the band structures of graphene and hBN around the $M$ point, and the similarity of their $\pi$ plasmons are by no means a coincidence.
Actually, similar to what happens in graphite \cite{Marinopoulos/etal:2004}, the $\pi$ plasmons stem from the collective electronic interband excitations
around the $M$ point, which is a saddle point in the BZ, and contributes most strongly to the density of states (DOS). Furthermore, in contrast to 3D materials,
the excitation energy approaches to the value
of single-particle energy gap as $q \rightarrow 0$ for plasmon excitations in 2D materials originating from collective interband transition with a finite gap. 
This explains the blue shift of the $\pi$ plasmons in hBN compared to that in graphene.
Below we will examine separately the dispersion relations of Dirac plasmons and $\pi$ plasmons in more details.

\subsubsection{Dirac plasmons}
\begin{figure}[ht]
  \begin{picture}(200,410)(0,0)
     \put(100,275){\makebox(0,0){\includegraphics[width=0.4\textwidth,clip]{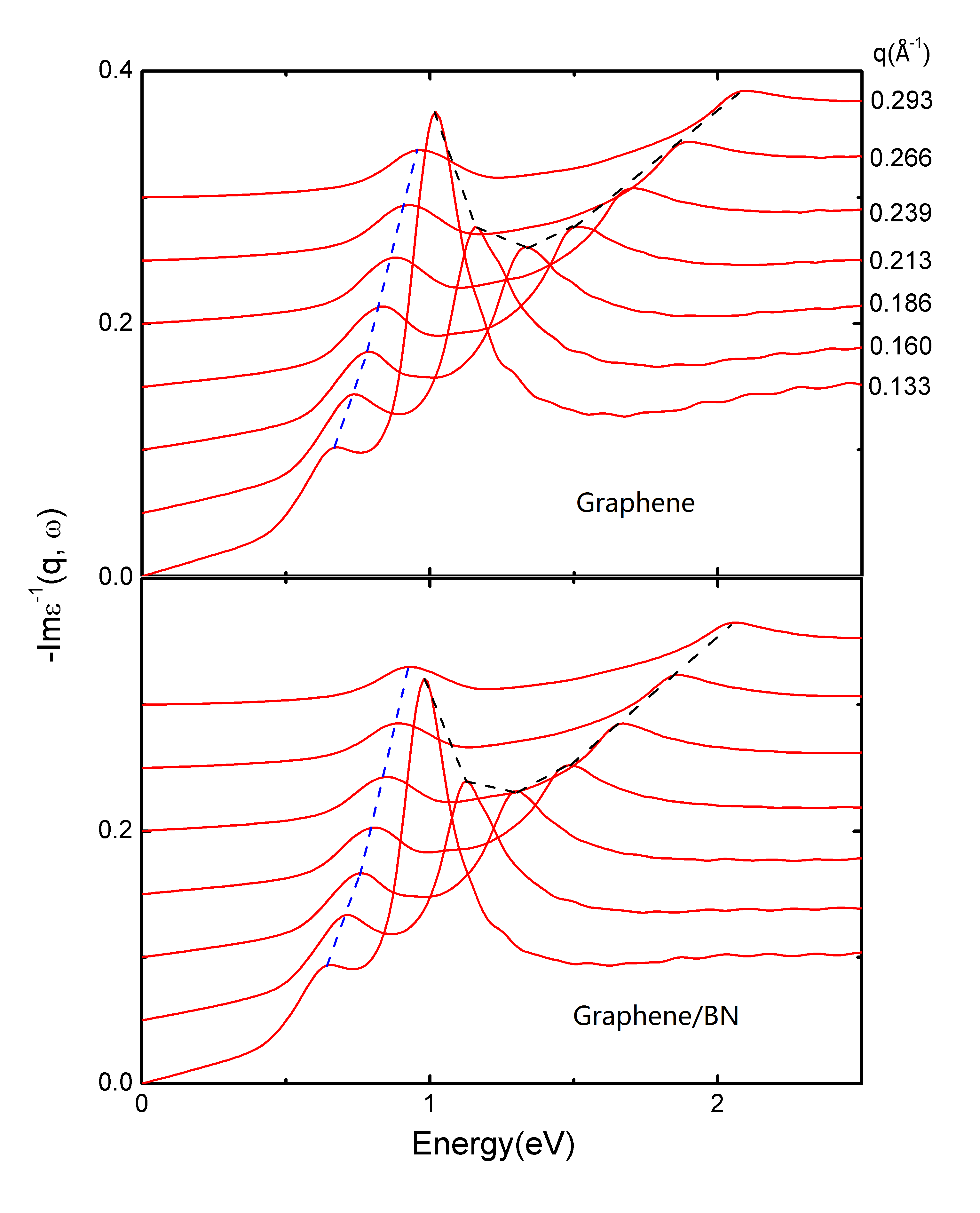}}}
     \put(100,65){\makebox(0,0){\includegraphics[width=0.45\textwidth,clip]{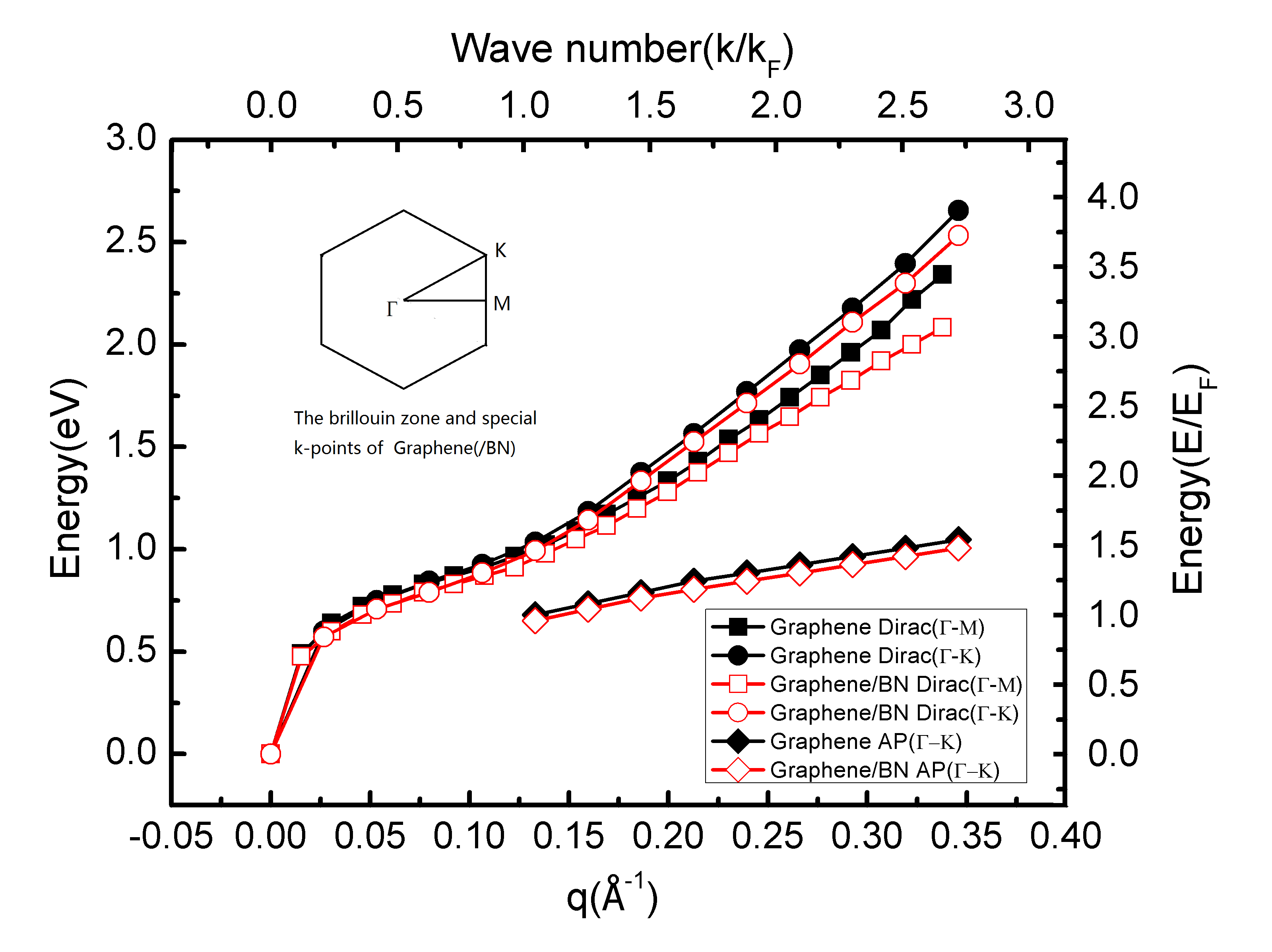}}}
     \put(20,400){\makebox(0,0){(a)}}
     \put(20,145){\makebox(0,0){(b)}}
  \end{picture}
\caption {(a): the loss spectra corresponding to Dirac plasmons along the $\Gamma-K$ direction in the BZ for doped graphene (upper panel)
and graphene/hBN (lower panel). (b): The Dispersion curves of the Dirac plasmons of along both $\Gamma-K$ and $\Gamma-M$ directions. The results
for the acoustic mode along the $\Gamma-K$ direction are also shown.}
\label{fig:Dirac_plasmon_dispersion}
\end{figure}

In Fig.~\ref{fig:Dirac_plasmon_dispersion}(a), the spectra of Dirac plasmons along the $\Gamma-K$ direction in doped graphene (upper panel) and graphene/hBN (lower panel) are presented.
Compared to the spectra along the $\Gamma-M$ direction, there is an additional shoulder peak along $\Gamma-K$, besides the ``conventional" 2D Dirac plasmons,  appearing at even lower energies.
This additional peak structure has been interpreted by Pisarra et al. \cite{Pisarra/etal:2014} as an acoustic plasmon mode. Specifically, plasmon excitations
along $\Gamma-K$
arise from oscillations of charge carriers with two different Fermi velocities. These two types of charge carrier oscillations can interact with each other, leading to an acoustic mode and
an optical mode \cite{Pines:1956}, with the latter being the conventional Dirac plasmons. Such an effect is not present for plasmons along the  $\Gamma-M$ direction. 
Comparing the spectra of graphene and graphene/hBN, we see that little has changed for both acoustic and conventional Dirac plasmons, upon including the hBN substrate.

The plasmon peak positions as a function of the momentum
transfer $q$ (the dispersion relations) are plotted in Fig.~ \ref{fig:Dirac_plasmon_dispersion}(b) for Dirac plasmons along both
the $\Gamma-M$ and $\Gamma-K$ directions. For the acoustic plasmon mode along the $\Gamma-K$ direction, the spectral weight gets vanishingly small
as $q\rightarrow 0$, making a precise determination of peak positions difficult. Hence the dispersion relation of the acoustic plasmons was not shown for
very small $q$'s in Fig.~\ref{fig:Dirac_plasmon_dispersion}(b).  For ``conventional" Dirac plasmons with $q<k_\text{F}$($\sim$ 0.12 \AA$^{-1}$), the dispersion
roughly follows a $\sqrt{q}$ behavior, in agreement with what was originally found in model studies \cite{Hwang/DasSarma:2007}. For large $q$'s, where
the Landau damping comes into play, the conventional Dirac plasmons actually display a quasi-linear dispersion behavior, as can be seen from
Fig.~\ref{fig:Dirac_plasmon_dispersion}(b).  Furthermore, both the anisotropic and substrate effects can be seen in Fig~\ref{fig:Dirac_plasmon_dispersion}(b).
Namely, for large $q$'s, the plasmon energies are larger along the $\Gamma-K$ direction compared to the $\Gamma-M$ direction for the same $q$, i.e.,
the $\Gamma-M$ dispersion curve is below that of $\Gamma-K$. Also, a slight red shift can be observed for large $q$'s, when the hBN substrate is included.


\subsubsection{$\pi$ plasmons}
\begin{figure}[ht]
\centering
\includegraphics[width=0.45\textwidth]{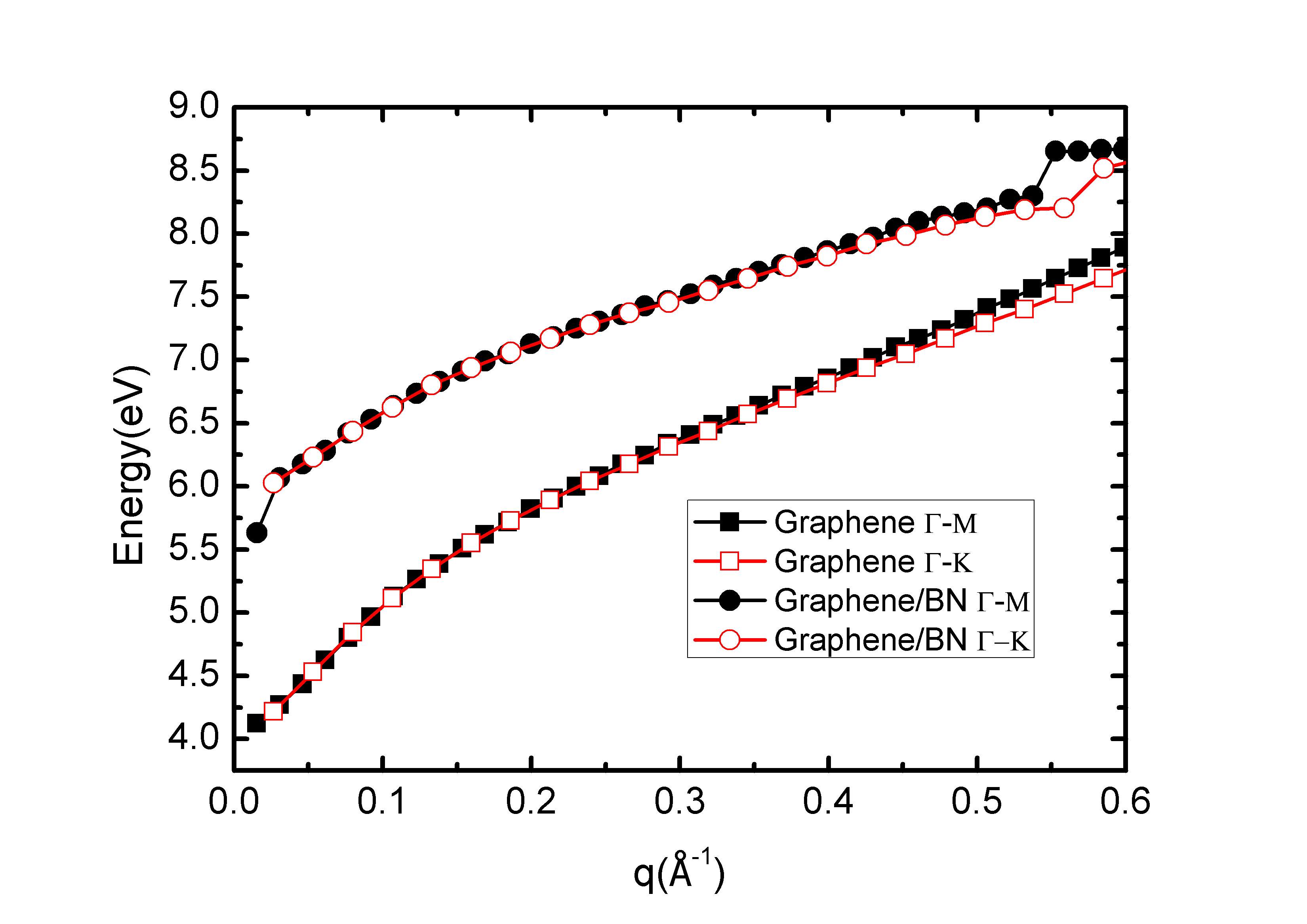}
\caption {The Dispersion behaviors of the $\pi$ plasmons of graphene (squares) and graphene/BN (circles) as a function
of the $\bfq$ points along both the $\Gamma-M$ (solid symbols) and $\Gamma-K$ (open symbols) directions. For graphene/BN,
the dispersion of the majority peak (cf. Fig.~\ref{fig:whole_spectra}) is plotted.}
\label{fig:Pi_plasmon_dispersion}
\end{figure}

Next we examine the dispersion behavior of the $\pi$ plasmons.
In Fig~\ref{fig:Pi_plasmon_dispersion}, dispersion relations of $\pi$ plasmons in graphene and graphene/hBN are presented, along both $\Gamma-K$ and
$\Gamma-M$ directions. For graphene/hBN, as discussed above, the plasmons in the $\pi$ region are split into several sub-peaks, and here only the majority
sub-peak, arising mainly from hBN, is plotted. The dispersion behavior of graphene/hBN $\pi$ plasmons gets a bit complex for momentum transfer $q >=0.55$ \AA$^{-1}$, 
where kinks can be seen in the dispersion curve. This is due to the fact that the hybridization effect gets strong for large $q$'s, associated with substantial
spectral weight transfer among the sub-peaks. Consequently, it is not possible any more to identity the original dominating (hBN-originated) peak. Also, 
for $q> 0.5$ \AA$^{-1}$, the dispersion along $\Gamma-K$ and $\Gamma-M$ start to deviate from each other, and an anisotropy effect becomes visible.

In this connection, we would like to stress that the dispersion relation of $\pi$ plasmons in graphene is still a debated issue. Except perhaps for
one work \footnote{The dispersion behavior of $\pi$ plasmons for small $\bfq$'s was not explicitly discussed in Ref.~[\onlinecite{Huang/Lin/Chuu:1997}], but the plot
in Fig 4 of this paper suggests a parabolic behavior, as seen by eyes.}, most earlier theoretical and  experimental studies \cite{Kramberger/etal:2008,Lu/etal:2009,Yan/etal:2011,Kinyanjui/etal:2012}
reported a quasi-linear dispersion behavior for $\pi$ plasmons. This generally accepted view was recently challenged by Liou \textit{et al.} \cite{Liou/etal:2015},
who proposed a $\sqrt{q}$ dispersion relation based on their high-momentum-resolution EELS experiment. From Fig.~\ref{fig:Pi_plasmon_dispersion}, however, one can
see that our first-principles TDDFT-RPA result cannot be fitted for by a simple linear or $\sqrt{q}$ model. This inconsistency motivated us to analyze this issue
more deeply in terms of a tight-binding model involving only the $\pi$ and $\pi*$ bands of graphene. Our model analysis suggest that the theoretically more
appropriate dispersion behavior of $\pi$ plasmons for small $q$'s should be  $\sqrt{E_{g,M}^2 + \beta q}$, where $E_{g,M}$ is the energy gap of graphene at 
the $M$ point.
Further details of this derivation are given in Appendix~\ref{app:pi_plasmon}. The essence behind our derivation is to recognize that the $\pi$ plasmons in
graphene stem from the collective excitations from the $\pi$ to $\pi^\ast$ band around the $M$ point. The $M$ point in the BZ represents a saddle point of
the $\pi$/$\pi^\ast$ bands, yielding a van Hove singularity in the DOS at energy $E=E_{g,M}$.
Thus the collective interband transition in the vicinity of the $M$ point dominates the contributions
to $\pi$ plasmons at small $q$'s.  For a 2D system like graphene, the plasmon excitation energy of this type approaches the single-particle energy gap
at the $M$ point as $q$ goes to 0.

\begin{figure}[ht]
\centering
\includegraphics[width=0.45\textwidth]{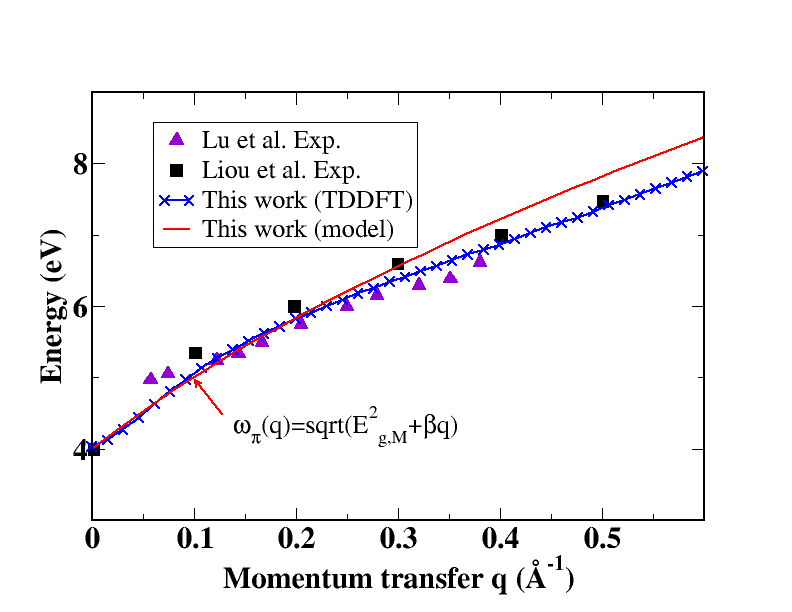}
\caption{The model dispersion curve of $\pi$ plasmons $\omega_\pi (q) = \sqrt{E_{g,M}^2 + \beta q}$ with $E_{g,M}=4$ eV and $\beta= 90$ eV$^2\cdot$\AA, in
 comparison with the dispersion curves obtained by TDDFT-RPA, and the experimental results of Liou \textit{et al.}
 [\onlinecite{Liou/etal:2015}] and Lu \textit{et al.} [\onlinecite{Lu/etal:2009}]. }
\label{fig:Pi_plasmon_comp}
\end{figure}

In Fig~\ref{fig:Pi_plasmon_comp} we plotted the model dispersion curve of $\omega_{\pi}(q) = \sqrt{E_{g,M}^2 + \beta q }$ with $\beta = 90$~ eV$^2\cdot$\AA,
and the first-principles TDDFT-RPA dispersion curve along $\Gamma-M$, which was already shown in Fig.~\ref{fig:Pi_plasmon_dispersion}. The value of the parameter
was chosen to fit well the slope of the first-principles results in the small $\bfq$'s region.  Also presented for comparison
are the experimental data by Liou \textit{et al.} \cite{Liou/etal:2015} and Lu \textit{et al.} \cite{Lu/etal:2009}. It can be seen that, with
one adjustable parameter, the model curve agrees very well with the first-principles one for $q<0.3$~\AA$^{-1}$, which is
the regime that the model is expected to be valid.  The overall agreement with the experimental data is satisfactory, though some remaining discrepancy is
noticeable. One may notice that the experimental data from the two groups also show some scattering, and especially the data for 0.01  $<q<$ 0.1 \AA$^{-1}$ 
are missing, which are important for an unambiguous determination for the $q\rightarrow 0$ behavior of the plasmon dispersion.
We note that the argument of Liou \textit{et al.} \cite{Liou/etal:2015} for the $\sqrt{q}$-dependence
behavior of $\pi$ plasmons was based on a 2-dimensionally (2D) free-electron gas model for graphene. Such a model will certainly be valid for describing the Dirac
plasmons in doped graphene. However, the $\pi$ plasmons stem from the collective excitations from the $\pi$ band to the $\pi^\ast$ around the $M$ point,
which has an energy gap of $\sim$4 eV. From this standpoint, we argue that the dispersion behavior of $\pi$ plasmons should follow that of a 2D insulator, 
instead of a 2D metal.  In this context, we call for more careful experiments to test our model.

\subsection{Lifetimes}

From the perspective of technological applications, the lifetime of plasmons is an important quantity to care about. Theoretically the lifetime of plasmon excitations is
inversely proportional to the width of the spectral peak, and hence the calculation of the lifetime is equivalent to determining the line width of the plasmon spectral peaks.
Figure~\ref{fig:whole_spectra} clearly shows that Dirac plasmons have very sharp peaks for small $q$'s; these peaks however quickly broadens and diminishes as $q$ increases.
This is consistent with the common understanding, as gained from model studies \cite{Hwang/DasSarma:2007,Wunsch/etal:2006}, that Dirac plasmons have infinite lifetimes within
RPA for momentum transfers below certain cutoff $q_c$. (For the doping level considered in this work ($5.1 \times 10^{13}$ cm$^{-2}$),
$q_c\sim 0.8 k_\text{F} =0.1$ \AA$^{-1}$.) Above $q_c$, the Landau damping kicks in, due to the merging of plasmon excitations and individual particle-hole excitations, and consequently the Dirac plasmons start to gain a finite lifetime.

However, an accurate determination of the lifetimes numerically in an \textit{ab initio} calculation is not entirely trivial. This is because
the actual width of the peaks
depends on the choice of the small positive parameter $\eta$ in Eqs.~(\ref{eq:chi_0}) and (\ref{Eq:chi_0_KK}).
By systematically reducing the value of $\eta$, one can in principle get more and more accurate peak widths, but care must be taken to to employ
denser $\bfk$ and frequency grids to get smooth and converged spectral peaks.
In this work, we adopt the following procedure to calculate the line width: first, the full widths at
half maximum (FWHM) of the plasmon peaks are determined for several different $\eta$ values ranging from 0.01 Ry to 0.001 Ry, and for each $\eta$, the spectrum is converged with respect to
the number of $\bfk$ and frequency points. Second, the FWHM's at finite $\eta$'s are extrapolated to the
limit of $\eta=0$. We consider the extrapolated FWHM as the final converged line-width at RPA level. Further details about our FWHM determination procedure 
is given in Appendix~\ref{app:FWHM}.
We successfully applied this procedure to accurately determine the FWHM's of Dirac and
$\pi$ plasmons of graphene.  However, we admit that this procedure does not go without
limitations. Especially, when there are several peaks very close to each other, an unambiguous determination of the width of each individual peak gets very
difficult. This is the case, e.g., for the $\pi$ plasmons in graphene/hBN, and $\pi$+$\sigma$ plasmons.

\begin{figure}[h]
\centering
\includegraphics[width=0.45\textwidth]{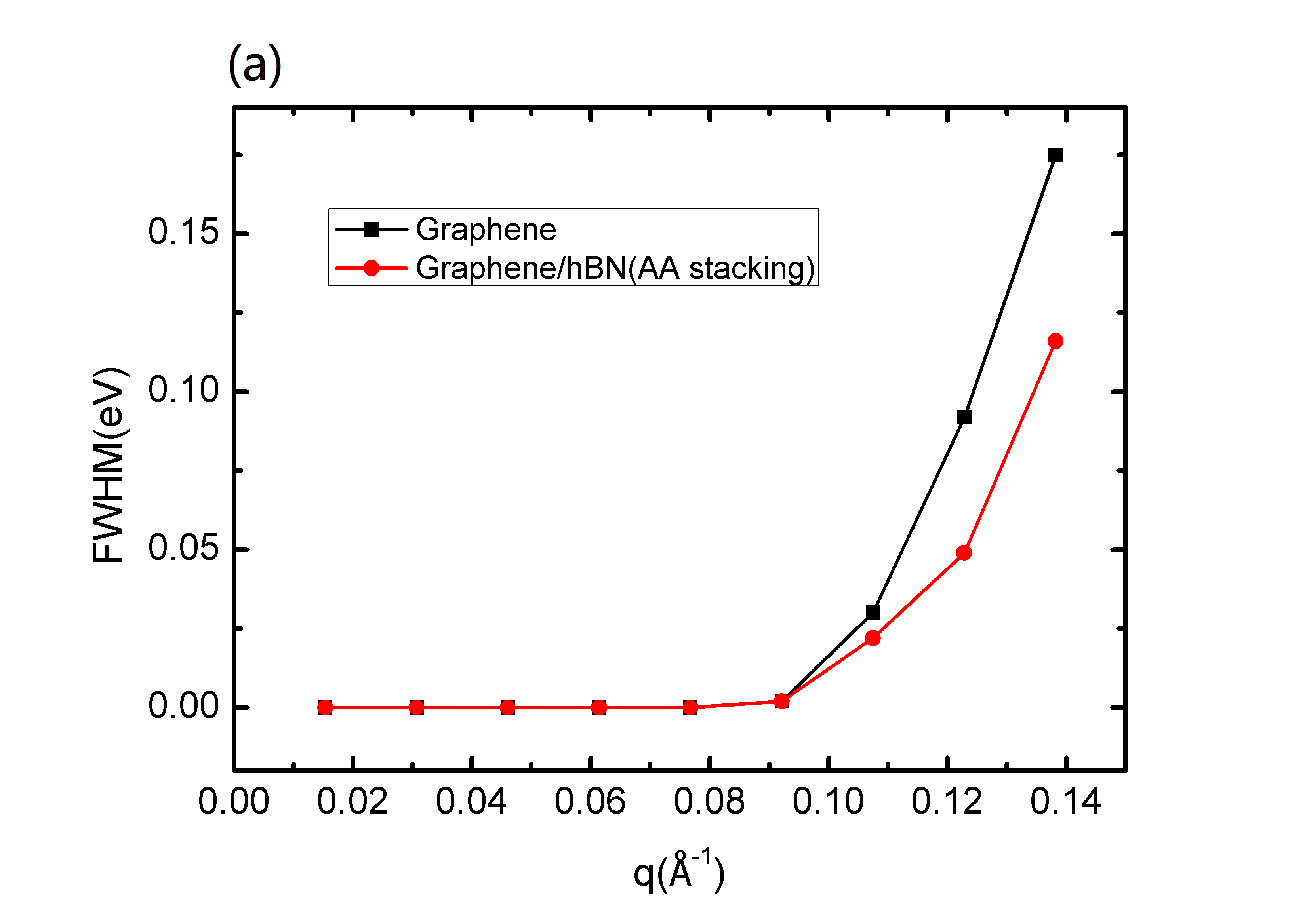}
\includegraphics[width=0.45\textwidth]{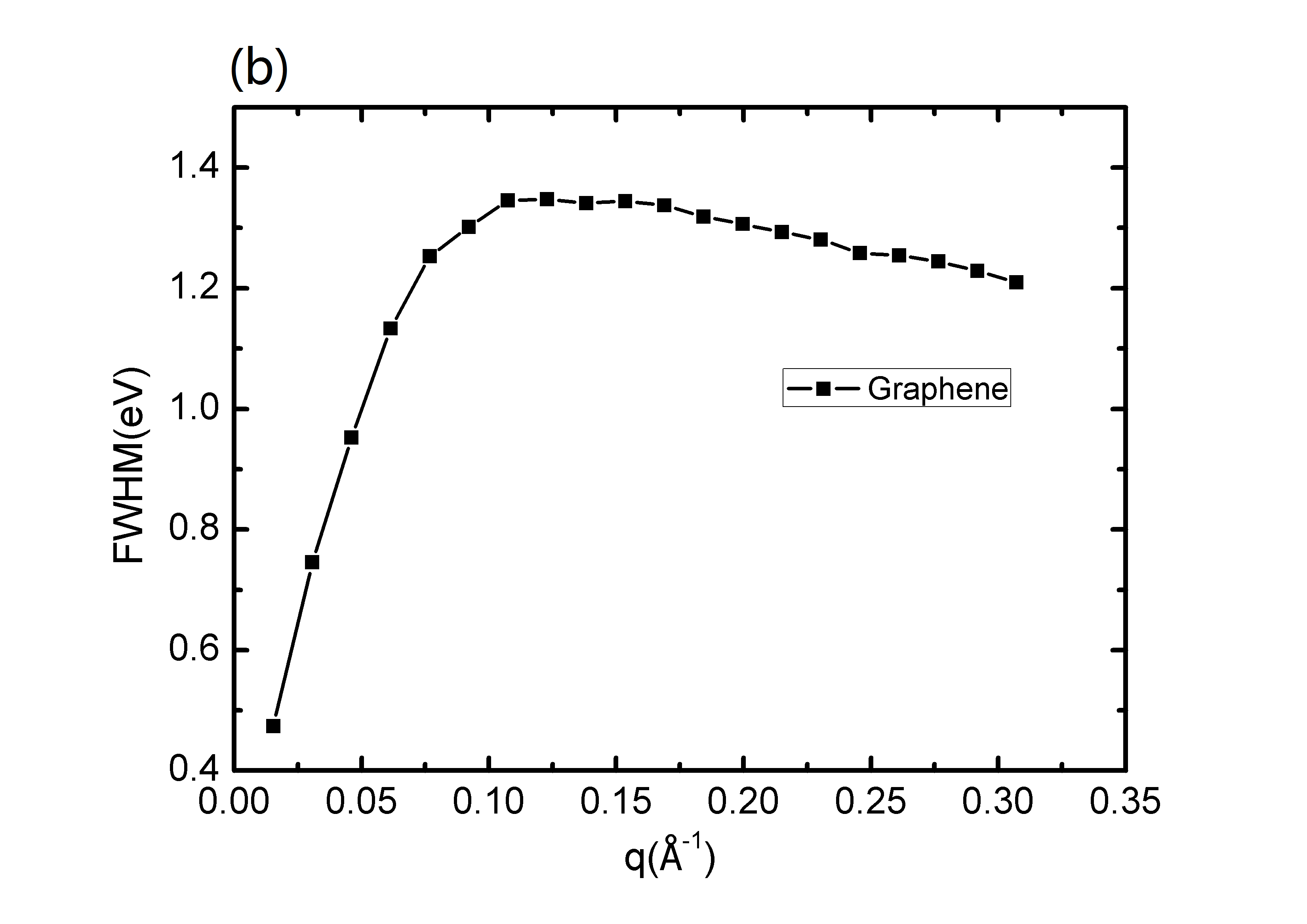}
\caption {The extrapolated ($\eta$ = 0) FWHM of Dirac (a) and $\pi$ (b) plasmons along the $\Gamma-M$ direction. For Dirac plasmon, results
are shown for both freestanding graphene and graphene/hBN. }
\label{fig:lifetime}
\end{figure}

In Fig~\ref{fig:lifetime} the extrapolated FWHM values for both Dirac and $\pi$ plasmons are presented as a function of the momentum transfer $q$. For Dirac
plasmons (Fig~\ref{fig:lifetime}(a)), the extrapolated FWHM value is strictly zero for $q < q_c$, indicating an infinite lifetime in this regime. For $q>q_c$, however, the FWHM of Dirac plasmons increases
quadratically as $(q-q_c)^2$.
This behavior for FWHM's of Dirac plasmons is in close agreement with the results of Wunsch \textit {et al.} \cite{Wunsch/etal:2006}, obtained from model studies.
Interestingly, in this case, the hBN substrate seems to have the tendency of reducing the line-width, i.e., increasing the lifetimes of Dirac plasmons $q>q_c$.
However, the effect here is purely electronic and different from the phonon and impurity scattering effects as discussed in
Ref.~[\onlinecite{Woessner/etal:2015}].

Distinct from Dirac plasmons, the $\pi$ and $\pi$+$\sigma$ plasmons have finite lifetimes for all momenta $\bfq$, because in these cases the Landau damping mechanism
is always active. Here only the FWHM values for $\pi$ plasmons in freestanding graphene are presented, since the above described numerical procedure for
determining FWHM does not yield reliable results for $\pi$ plasmons in graphene/hBN and for $\pi+\sigma$ plasmons. From Fig.~\ref{fig:lifetime}(b), one can see that the FWHM value of $\pi$ plasmons first steadily increase as a function of $q$, reaching its maximum for
$q\sim 0.12$ \AA$^{-1}$, and then, surprisingly, the line width starts to slowly decrease for even larger $q$'s. Such a non-monotonic behavior
for $\pi$ plasmon lifetimes in pure graphene is not yet well understood.

Finally, we emphasize again that the lifetime studies reported in this work is purely electronic at the RPA level, and hence the conclusion might not be directly applicable
to realistic situations. Yet this study is meaningful from a numerical point of view, since a reliable determination of plasmons line-widths from first-principles
calculations appears to be a challenging task. Our procedure could provide reference numbers for simple situations. Plasmon lifetime studies incorporating
effects from phonons, impurities, and disorders \cite{Jablan/etal:2009,Principi/etal:2014}, as well as electronic effects beyond RPA have been reported in
literature at model level\cite{Principi/etal:2013}.  Including these effects in a first-principles way is in principle also feasible, but goes beyond the scope of this work.

\section{\label{sec:conclusion}Summary}
Plasmon excitations in graphene and graphene-related materials have intriguing properties that hold great promises for technological applications. 
In this work, we developed a first-principles TDDFT-RPA module within the ABACUS software package \cite{abacusweb}, which allows for accurately simulating the 
plasmon excitations in graphene and graphene/hBN heterostructure for a large energy window (from 0 to $\sim 30$ eV). Regarding the controversial nature of 
the dispersion behavior of $\pi$ plasmons, our first-principles results and model analysis indicated that a theoretically more sound dispersion relation should be 
$\omega_\pi(q) = \sqrt{E_{g,M}^2+\beta q}$ at small $q$'s, in stark distinction from 
previous proposals. The essential physics behind this is that the $\pi$ plasmons in graphene arise from collective interband excitations that has a finite gap,
and hence the asymptotic behavior at $\bfq\rightarrow 0$ is different from that of Dirac plasmons, which come from collective gapless intraband transitions.
Finally, we demonstrated that, to extract accurate lifetime from the computed spectra, care must be taken to extrapolate the results to the limit of 
$\eta \rightarrow 0$, where $\eta$ is the technical broadening parameter used in dynamical response function calculations.

\begin{appendix}
\section{Comparison of graphene/h-BN heterostructure with different stackings.}
\label{app:g-hBN_stacking_comp}

\begin{figure}[ht]
\centering
\includegraphics[width=0.45\textwidth]{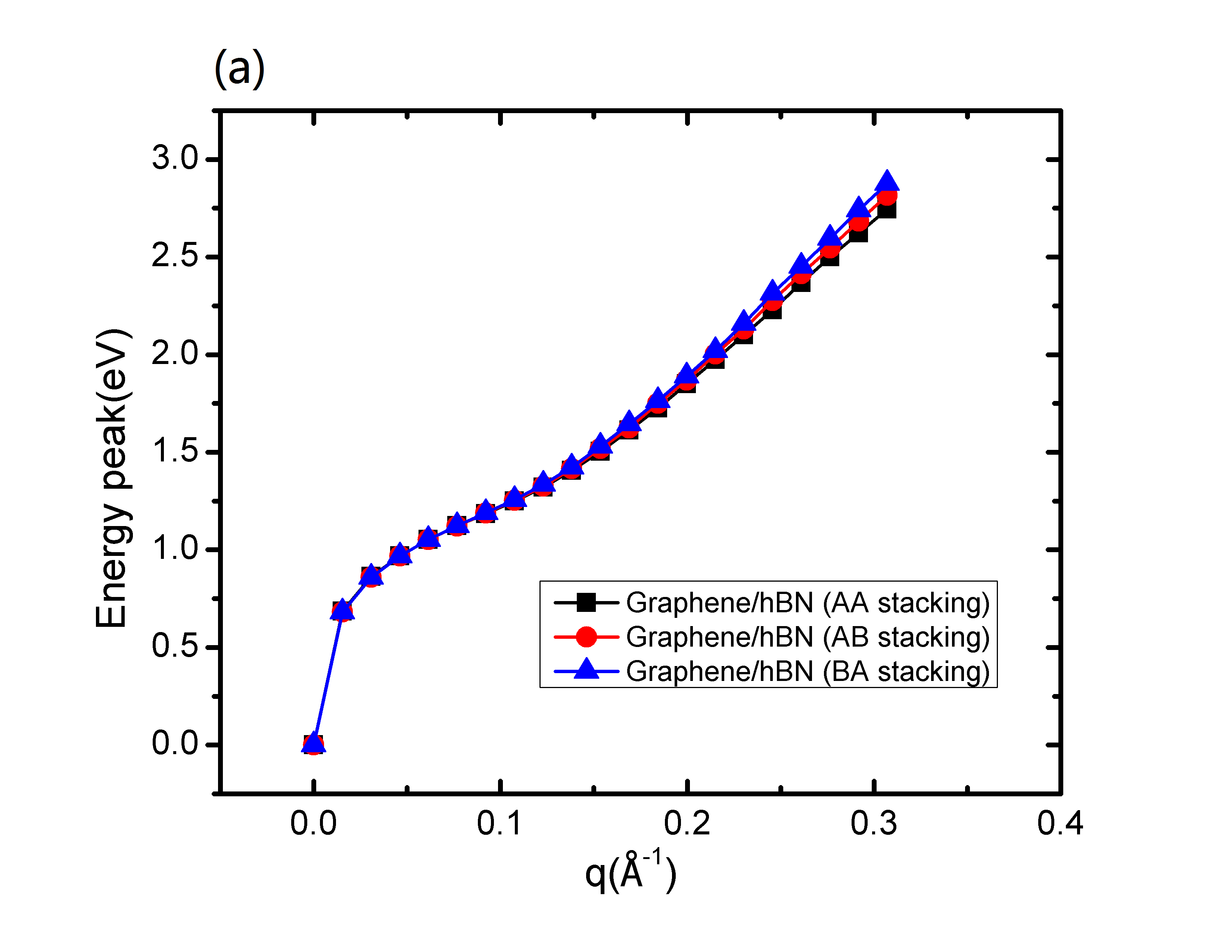}
\includegraphics[width=0.45\textwidth]{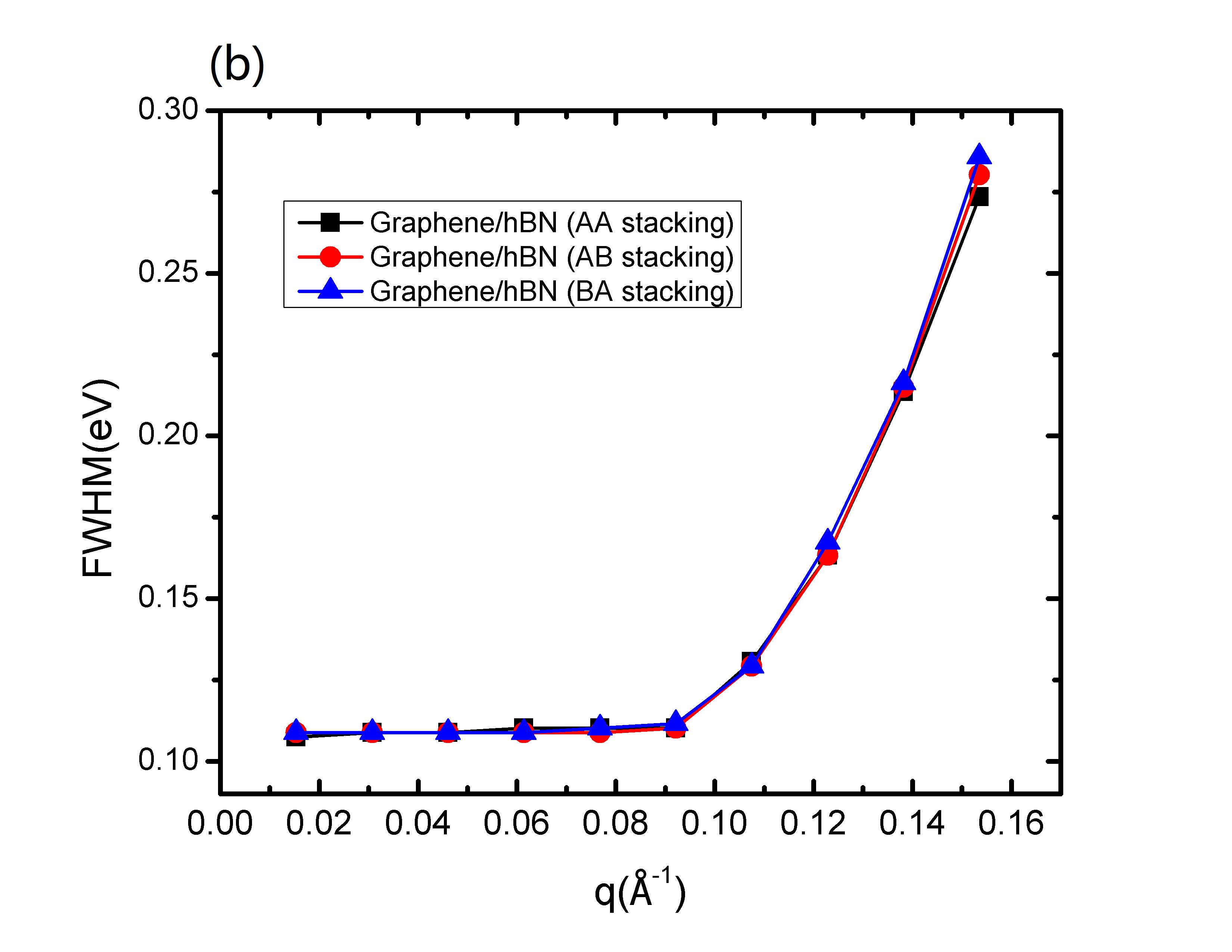}
\caption {(a) The dispersion relations of Dirac plasmons in the graphene/h-BN systems with three different (AA, AB, BA) stackings.
(b) the corresponding FWHM's of Dirac plasmons with the three different stackings. Results are obtained with $\eta$ = 0.004Ry (hence
the nonvanishing FWHM below $q_c$).}
\label{fig:g_hBN_stacking}                                                                                                                            \
\end{figure}

In the main text, for simplicity, we only presented the computational results of the graphene/h-BN heterostructure with AA stacking.
For completeness, here we also present in Fig.~\ref{fig:g_hBN_stacking} the results of dispersion relations and lifetimes (FWHM's of the peaks)
for the Dirac plasmons with other types of stackings. Figure.~\ref{fig:g_hBN_stacking} shows that the different stackings have little impact
on both the dispersion behavior and lifetime. We also did a similar comparison study for $\pi$ plasmons, and arrived at essentially the
same conclusion.

\section{On the dispersion of the $\pi$ plasmons in graphene}
\label{app:pi_plasmon}
We start with a tight-binding model description of graphene
   \begin{equation}
        H= t\sum_{\langle i,j \rangle} \hat{c}_{i}^\ast \hat{c}_{j} \,
   \end{equation}
where $t$ is the hopping integral between neighboring sites on a honeycomb lattice. The resultant dispersion of the $\pi$ (valence) and $\pi^\ast$
(conduction) bands reads
\begin{widetext}
 \begin{equation}
   \epsilon^{c,v} (k_x, k_y)=\pm t \left[1+4\text{cos}\left(\frac{3ak_y}{2}\right)
          \text{cos}\left(\frac{\sqrt{3}ak_x}{2}\right)+4\text{cos}^2\left(\frac{\sqrt{3}ak_x}{2}\right) \right]^{1/2}  \, ,
   \label{eq:TB_eigenvalue}
 \end{equation}
\end{widetext}
with $a$ being the distance between neighboring C atoms, and $\bfk=(k_x,k_y)$ are the two-dimensional wavevectors in the BZ.
The real part of the dielectric function of graphene given by such a tight-binding model, within the random-phase approximation, can be obtained as
 \begin{equation}
   {Re}\{\varepsilon(\bfq,\omega)\} = 1 -\frac{2\pi}{q} \sum_{\bfk} \frac{ |\langle v,\bfk | e^{-i\bfq\cdot \bfr} |c,\bfk+\bfq \rangle|^2
                                                           2(\epsilon_{c,\bfk+\bfq}-\epsilon_{v,\bfk})}
                                                          {\omega^2-(\epsilon_{c,\bfk+\bfq}-\epsilon_{v,\bfk})^2} \, ,
  \label{appeq:dielec_func}
 \end{equation}
where $\epsilon_{v,\bfk}$/$\epsilon_{c,\bfk}$ are the valence/conduction band energies as given by Eq.~(\ref{eq:TB_eigenvalue}),
and $|v,\bfk\rangle$/$|c,\bfk\rangle$ are the corresponding eigenvectors.
We note that that $(k_x=0,k_y=2\pi/3a)$ corresponds to one $M$ point in the BZ. The excitation energy gap in the vicinity of this $M$ point
can be obtained by a Taylor expansion,
  \begin{widetext}
    \begin{align}
       E_g(\bfk',\bfq) \approx & t\left[1+\frac{9}{2}a^2{\left(k'_y+q_y\right)}^{2} -\frac{3}{2}a^2{\left(k'_x+q_x\right)}^{2} \right]^{1/2} +
                                t\left[1+\frac{9}{2}a^2{k'_y}^{2} -\frac{3}{2}a^2{k'_x}^{2} \right]^{1/2} \nonumber \\
                     \approx & E_{g,M} + \frac{9}{4}t a^2 \left(2{k'_y}^{2} + 2k'_yq_y +q_y^2 \right) - \frac{3}{4}t a^2\left(2{k'_x}^{2} + 2k'_xq_x + q_x^2 \right)                                 \nonumber \\
                        = & E_{g,M} + \Delta E_g(\bfk',\bfq)
         \label{eq:gap_M} \, ,
    \end{align}
  \end{widetext}
where $\bfk'=(k'_x,k'_y)=(k_x, k_y-2\pi/3a)$ is the coordinate of a $\bfk$ point with reference to the $M$ point, and
$E_{g,M}$ is the direct band gap at the $M$ point. From Eq.~(\ref{eq:gap_M}) one may realize that the $M$ point is a saddle point
in the band dispersion of graphene, and $E_{g,M}$ represents a van Hove singularity in the single-particle density of states (DOS)
of graphene.

For $q\rightarrow 0$, the oscillator strength in the numerator of Eq.~(\ref{appeq:dielec_func}) becomes
 \begin{align}
 \text{lim}_{q\rightarrow 0} \langle v,\bfk | e^{-i\bfq\cdot \bfr} |c,\bfk+\bfq \rangle = & -iq \frac{\hbar^2}{m_e}\frac{\langle v,\bfk|\nabla_\bfr|c,\bfk\rangle} {
\epsilon_{c}(\bfk)-\epsilon_{v}(\bfk)} \nonumber  \\
   = & -iq \frac{\hbar^2}{m_e} \frac{p(\bfk)} {\epsilon_{c}(\bfk)-\epsilon_{v}(\bfk)} \, .
 \end{align}
where $p(\bfk)= \langle v,\bfk|\nabla_\bfr|c,\bfk\rangle$, and $m_e$ is the mass of the electron.
Furthermore, we are interested in the $\pi$ plasmons which originate from the collective $\pi\rightarrow\pi^\ast$ interband excitations
around the $M$ point. Therefore the $k$-integration in Eq~(\ref{appeq:dielec_func}) can be restricted to the vicinity of $M$ (i.e.
by setting $|\bfk'|<=k_{cut}$), and consequently the dielectric function at $q\rightarrow 0$ is simplified to
 \begin{widetext}
 \begin{align}
  \displaystyle
  {Re}\{\varepsilon(\bfq,\omega)\}& = 1 -\frac{2\pi}{q}q^2 \frac{\hbar^4}{m_e^2} \sum_{|\bfk'|<=k_{cut}} \frac{ p^2(\bfk')2E_g(\bfk',\bfq) }
                         {\left[\omega^2-\left(E_{g,M}+\Delta E_g(\bfk',\bfq)\right)^2\right] \left(\epsilon_{c}(\bfk')-\epsilon_{v}(\bfk')\right)^2} \,  \nonumber \\
                      & \approx \frac{q}{\omega^2-E_{g,M}^2} \frac{2\pi \hbar^4}{m_e^2} \sum_{|\bfk'|<=k_{cut}}  \frac{ p^2(\bfk')2E_g(\bfk',\bfq) }
                         {\left[1-\frac{2E_{g,M}E_g(\bfk',\bfq)}{\omega^2-E_{g,M}^2}\right] E_g^2(\bfk',0)} \nonumber \\
                          & = 1- \frac{q\beta(\bfq,\omega)}{\omega^2-E_{g,M}^2} \, .
  \label{appeq:dielec_func_simple}
 \end{align}
 \end{widetext}
The precise form of the parameter $\beta$ introduced in Eq.~(\ref{appeq:dielec_func_simple}) in principle depends on $\bfq$ and $\omega$,
but the dependence is of higher order. To the zeroth order approximation, $\beta$ can be taken to be a constant, which is valid for small $q$'s. Also, since
a proper value  for the cutting wavevector $k_{cut}$ is not known \textit{a priori},  we cannot determine $\beta$ reliably within this model.
Therefore, in this work $\beta$ is treated as a fitting parameter.

The plasmon dispersion relation can be determined by searching for the zeros of the dielectric function ${Re}\{\varepsilon(\bfq,\omega)\}=0$. Hence we end up with,
  \begin{equation}
       1- \frac{\beta q}{\omega^2-E_{g,M}^2}  = 0
  \end{equation}
or
  \begin{equation}
          \omega(q) = \sqrt{E_{g,M}^2 + \beta q} \, .
  \end{equation}




\section{Procedure to determine FWHM}
\label{app:FWHM}
In this section, we describe in further details how the plasmons lifetimes (or more precisely the FWHM's) are determined in our work.
First, the loss spectra were calculated for several different $\eta$ parameters.  For a given $\eta$, the loss spectra calculations
are converged with other technical parameters such as the number of $\bfk$-point sampling in BZ, and the number of frequency grid points.
 The position and height ($h$) of the spectral peak center are then determined, together with
the positions of half height ($h/2$) on the left- and right-hand sides of the peak center, denoted as $\omega_l$ and $\omega_r$ respectively.
The FWHM for a given $\eta$ and $\bfq$ is then given by $\omega_r-\omega_l$. The final FWHM value is then obtained by extrapolating
the FWHM values for finite $\eta$'s to $\eta=0$.

\begin{figure}[ht]
\centering
\includegraphics[width=0.4\textwidth]{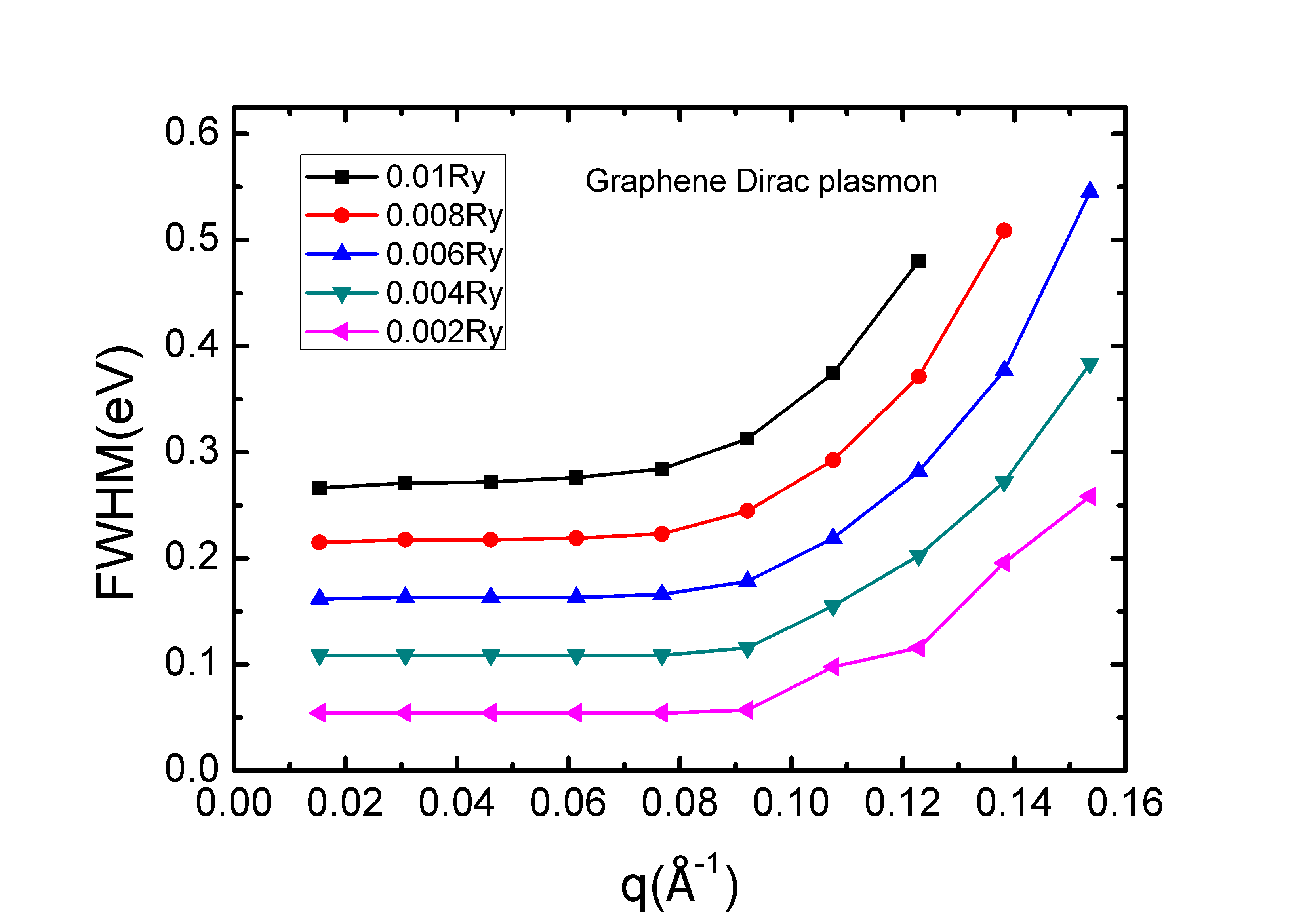}
\caption {The FWHM's of the conventional Dirac plasmon peaks as a function of $q$ along the $\Gamma-M$ direction for different
 $\eta$ values from 0.002 to 0.01 Ry.}
\label{fig:app_FWHM_etaqdep}
\end{figure}

Figure~\ref{fig:app_FWHM_etaqdep} shows the FWHM values of the conventional Dirac plasmon peaks in graphene for different $\eta$ values.
One can clearly see that the FWHM value depends appreciably on the parameter $\eta$. For $\eta=0.01$ Rydberg, a typical default choice
for first-principle TDDFT-RPA calculations, the FWHM value is overestimated by more than 0.25 eV for small $q$'s. As $\eta$ is reduced,
the FWHM value decreases steadily to zero. From Fig.~\ref{fig:app_FWHM_etaqdep} one can also see their is a critical momentum value $q_c$,
below which the FWHM value approaches all the way to zero as $\eta$ goes to zero.

\begin{figure}[ht]
\centering
\includegraphics[width=0.4\textwidth]{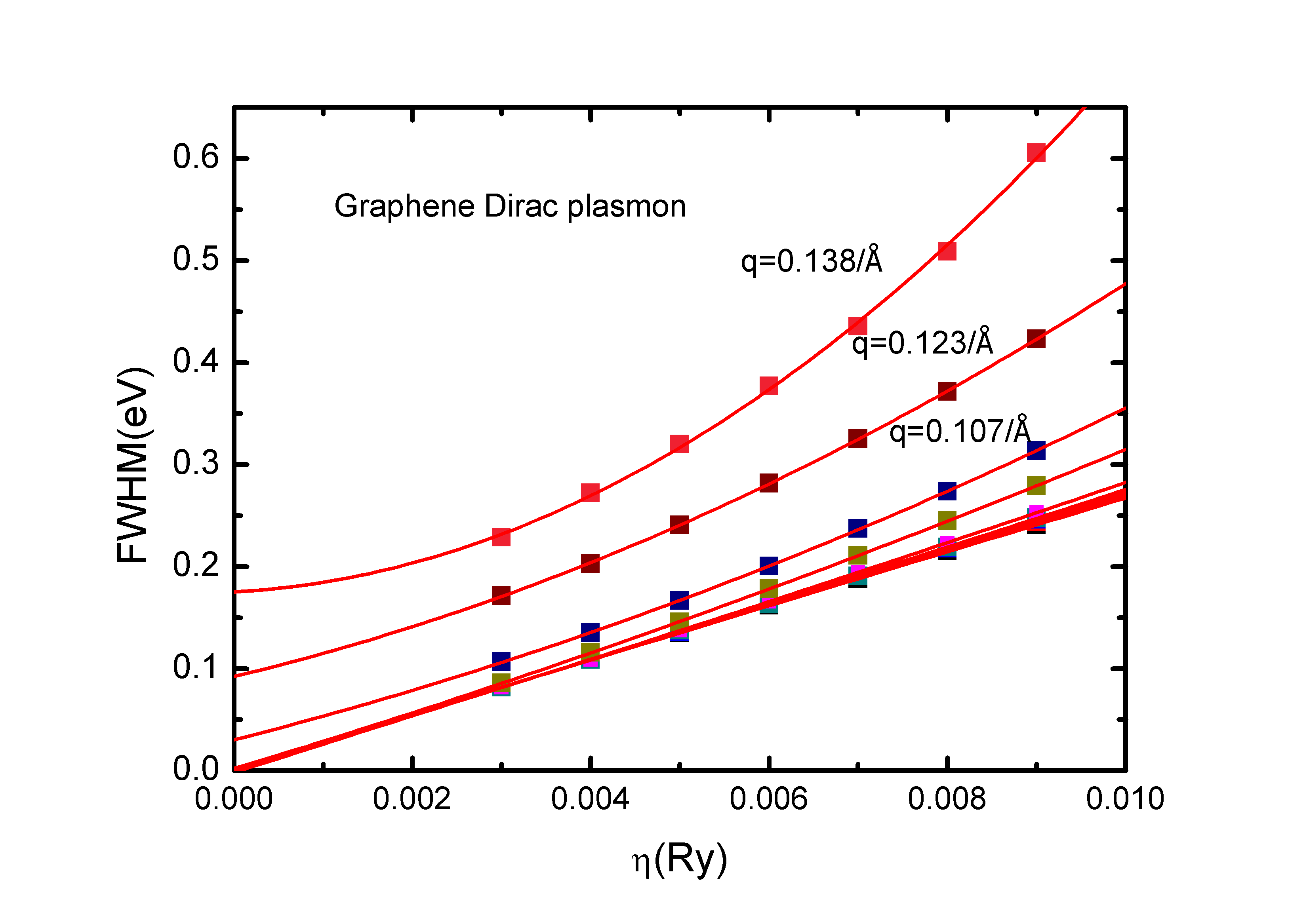}
\caption {The extrapolation of the FWHM value of the Dirac plasmons in graphene to the limit of $\eta$ = 0. For
$q <q_c 0.107$ $ \AA^{-1}$ ($\sim$ 0.8$k_F$), the FWHM is linearly proportional to $\eta$; for $q> q_c$,
the FWHM goes to a finite value quadratically as $\eta \rightarrow 0$.  }
\label{fig:app_FWHM_eta-extra}
\end{figure}

Figure~\ref{fig:app_FWHM_eta-extra} illustrates how the FWHM value is extrapolated to $\eta=0$ limit. Below the critical $q$ value
$q_c = 0.107 \AA^{-1}$, the actual FWHM value is linear proportional to $\eta$ (actually roughly $2\eta$). Thus the FWHM naturally
goes to zero at the $\eta=0$ limit. For $q>q_c$, the $\eta$-dependence of FWHM follows nicely a quadratic behavior, saturating at
a finite value at $\eta=0$. The thus obtained $\eta=0$ limit of FWHM for Dirac plasmons were presented in Fig.~\ref{fig:lifetime}.
Also presented are the FWHM results for $\pi$ plasmons, obtained using a similar procedure. However, for $\pi$ plasmons the FWHM
is finite for all $q$ values.

\end{appendix}
\section*{Acknowledgments}
The work was supported by the National Key Research and Development Program of China (Grants No. 2016YFB0201202). XR acknowledges the support
from Chinese National Science Foundation (Grant number 11374276, 11574283).
LH acknowledges the support from Chinese National Science Foundation (Grant number 11374275). The numerical calculations have
been done on the USTC HPC facilities.

\bibliography{./CommonBib}

\end{document}

%% file: newcommands.tex
\newcommand{\bfr}{ {\bf r}} 
\newcommand{\bfq}{ {\bf q}} 
\newcommand{\bfk}{ {\bf k}} 
\newcommand{\bfG}{ {\bf G}} 
\newcommand{\bfGp}{ {\bf G'}} 



%% file: plasmon_main.bbl
\begin{thebibliography}{41}
\expandafter\ifx\csname natexlab\endcsname\relax\def\natexlab#1{#1}\fi
\expandafter\ifx\csname bibnamefont\endcsname\relax
  \def\bibnamefont#1{#1}\fi
\expandafter\ifx\csname bibfnamefont\endcsname\relax
  \def\bibfnamefont#1{#1}\fi
\expandafter\ifx\csname citenamefont\endcsname\relax
  \def\citenamefont#1{#1}\fi
\expandafter\ifx\csname url\endcsname\relax
  \def\url#1{\texttt{#1}}\fi
\expandafter\ifx\csname urlprefix\endcsname\relax\def\urlprefix{URL }\fi
\providecommand{\bibinfo}[2]{#2}
\providecommand{\eprint}[2][]{\url{#2}}

\bibitem[{\citenamefont{Huang et~al.}(1997)\citenamefont{Huang, Lin, and
  Chuu}}]{Huang/Lin/Chuu:1997}
\bibinfo{author}{\bibfnamefont{C.~S.} \bibnamefont{Huang}},
  \bibinfo{author}{\bibfnamefont{M.~F.} \bibnamefont{Lin}}, \bibnamefont{and}
  \bibinfo{author}{\bibfnamefont{D.~S.} \bibnamefont{Chuu}},
  \bibinfo{journal}{Solid State Commun.} \textbf{\bibinfo{volume}{103}},
  \bibinfo{pages}{603} (\bibinfo{year}{1997}).

\bibitem[{\citenamefont{Shyu and Lin}(2000)}]{Shyu/Lin:2000}
\bibinfo{author}{\bibfnamefont{F.~L.} \bibnamefont{Shyu}} \bibnamefont{and}
  \bibinfo{author}{\bibfnamefont{M.~F.} \bibnamefont{Lin}},
  \bibinfo{journal}{Phys. Rev. B} \textbf{\bibinfo{volume}{62}},
  \bibinfo{pages}{8508} (\bibinfo{year}{2000}).

\bibitem[{\citenamefont{Wunsch et~al.}(2006)\citenamefont{Wunsch, Stauber,
  Sols, and Guinea}}]{Wunsch/etal:2006}
\bibinfo{author}{\bibfnamefont{B.}~\bibnamefont{Wunsch}},
  \bibinfo{author}{\bibfnamefont{T.}~\bibnamefont{Stauber}},
  \bibinfo{author}{\bibfnamefont{F.}~\bibnamefont{Sols}}, \bibnamefont{and}
  \bibinfo{author}{\bibfnamefont{F.}~\bibnamefont{Guinea}},
  \bibinfo{journal}{New J. Phys.} \textbf{\bibinfo{volume}{8}},
  \bibinfo{pages}{318} (\bibinfo{year}{2006}).

\bibitem[{\citenamefont{Hwang and {Das Sarma}}(2007)}]{Hwang/DasSarma:2007}
\bibinfo{author}{\bibfnamefont{E.~H.} \bibnamefont{Hwang}} \bibnamefont{and}
  \bibinfo{author}{\bibfnamefont{S.}~\bibnamefont{{Das Sarma}}},
  \bibinfo{journal}{Phys. Rev. B} \textbf{\bibinfo{volume}{75}},
  \bibinfo{pages}{205418} (\bibinfo{year}{2007}).

\bibitem[{\citenamefont{Jablan et~al.}(2009)\citenamefont{Jablan, Buljan, and
  Solja\v{c}i\'{c}}}]{Jablan/etal:2009}
\bibinfo{author}{\bibfnamefont{M.}~\bibnamefont{Jablan}},
  \bibinfo{author}{\bibfnamefont{H.}~\bibnamefont{Buljan}}, \bibnamefont{and}
  \bibinfo{author}{\bibfnamefont{M.}~\bibnamefont{Solja\v{c}i\'{c}}},
  \bibinfo{journal}{Phys. Rev. B} \textbf{\bibinfo{volume}{80}},
  \bibinfo{pages}{245435} (\bibinfo{year}{2009}).

\bibitem[{\citenamefont{Yan et~al.}(2011)\citenamefont{Yan, Thygesen, and
  Jacobsen}}]{Yan/etal:2011}
\bibinfo{author}{\bibfnamefont{J.}~\bibnamefont{Yan}},
  \bibinfo{author}{\bibfnamefont{K.~S.} \bibnamefont{Thygesen}},
  \bibnamefont{and} \bibinfo{author}{\bibfnamefont{K.~W.}
  \bibnamefont{Jacobsen}}, \bibinfo{journal}{Phys. Rev. Lett.}
  \textbf{\bibinfo{volume}{106}}, \bibinfo{pages}{146803}
  (\bibinfo{year}{2011}).

\bibitem[{\citenamefont{Eberlein et~al.}(2008)\citenamefont{Eberlein, Bangert,
  Nair, Jones, Gass, Bleloch, Novoselov, Geim, and
  Briddon}}]{Eberlein/etal:2008}
\bibinfo{author}{\bibfnamefont{T.}~\bibnamefont{Eberlein}},
  \bibinfo{author}{\bibfnamefont{U.}~\bibnamefont{Bangert}},
  \bibinfo{author}{\bibfnamefont{R.~R.} \bibnamefont{Nair}},
  \bibinfo{author}{\bibfnamefont{R.}~\bibnamefont{Jones}},
  \bibinfo{author}{\bibfnamefont{M.}~\bibnamefont{Gass}},
  \bibinfo{author}{\bibfnamefont{A.~L.} \bibnamefont{Bleloch}},
  \bibinfo{author}{\bibfnamefont{K.~S.} \bibnamefont{Novoselov}},
  \bibinfo{author}{\bibfnamefont{A.}~\bibnamefont{Geim}}, \bibnamefont{and}
  \bibinfo{author}{\bibfnamefont{P.~R.} \bibnamefont{Briddon}},
  \bibinfo{journal}{Phys. Rev. B} \textbf{\bibinfo{volume}{77}},
  \bibinfo{pages}{233406} (\bibinfo{year}{2008}).

\bibitem[{\citenamefont{Kramberger et~al.}(2008)\citenamefont{Kramberger,
  Hambach, Giorgetti, R{\"u}mmeli, Knupfer, Fink, B\"{u}chner, Reining,
  Einarsson, Maruyama et~al.}}]{Kramberger/etal:2008}
\bibinfo{author}{\bibfnamefont{C.}~\bibnamefont{Kramberger}},
  \bibinfo{author}{\bibfnamefont{R.}~\bibnamefont{Hambach}},
  \bibinfo{author}{\bibfnamefont{C.}~\bibnamefont{Giorgetti}},
  \bibinfo{author}{\bibfnamefont{M.~H.} \bibnamefont{R{\"u}mmeli}},
  \bibinfo{author}{\bibfnamefont{M.}~\bibnamefont{Knupfer}},
  \bibinfo{author}{\bibfnamefont{J.}~\bibnamefont{Fink}},
  \bibinfo{author}{\bibfnamefont{B.}~\bibnamefont{B\"{u}chner}},
  \bibinfo{author}{\bibfnamefont{L.}~\bibnamefont{Reining}},
  \bibinfo{author}{\bibfnamefont{E.}~\bibnamefont{Einarsson}},
  \bibinfo{author}{\bibfnamefont{S.}~\bibnamefont{Maruyama}},
  \bibnamefont{et~al.}, \bibinfo{journal}{Phys. Rev. Lett.}
  \textbf{\bibinfo{volume}{100}}, \bibinfo{pages}{196803}
  (\bibinfo{year}{2008}).

\bibitem[{\citenamefont{Liu et~al.}(2008)\citenamefont{Liu, Willis, Emtsev, and
  Seyller}}]{Liu/etal:2008}
\bibinfo{author}{\bibfnamefont{Y.}~\bibnamefont{Liu}},
  \bibinfo{author}{\bibfnamefont{R.~F.} \bibnamefont{Willis}},
  \bibinfo{author}{\bibfnamefont{K.~V.} \bibnamefont{Emtsev}},
  \bibnamefont{and} \bibinfo{author}{\bibfnamefont{T.}~\bibnamefont{Seyller}},
  \bibinfo{journal}{Phys. Rev. B} \textbf{\bibinfo{volume}{78}},
  \bibinfo{pages}{201403(R)} (\bibinfo{year}{2008}).

\bibitem[{\citenamefont{Lu et~al.}(2009)\citenamefont{Lu, Loh, Huang, Chen, and
  Wee}}]{Lu/etal:2009}
\bibinfo{author}{\bibfnamefont{J.}~\bibnamefont{Lu}},
  \bibinfo{author}{\bibfnamefont{K.~P.} \bibnamefont{Loh}},
  \bibinfo{author}{\bibfnamefont{H.}~\bibnamefont{Huang}},
  \bibinfo{author}{\bibfnamefont{W.}~\bibnamefont{Chen}}, \bibnamefont{and}
  \bibinfo{author}{\bibfnamefont{A.~T.~S.} \bibnamefont{Wee}},
  \bibinfo{journal}{Phys. Rev. B} \textbf{\bibinfo{volume}{80}},
  \bibinfo{pages}{113410} (\bibinfo{year}{2009}).

\bibitem[{\citenamefont{Koppens et~al.}(2011)\citenamefont{Koppens, Chang, ,
  and {de Abajo}}}]{Koppens/etal:2011}
\bibinfo{author}{\bibfnamefont{F.~H.~L.} \bibnamefont{Koppens}},
  \bibinfo{author}{\bibfnamefont{D.~E.} \bibnamefont{Chang}}, ,
  \bibnamefont{and} \bibinfo{author}{\bibfnamefont{F.~J.~G.} \bibnamefont{{de
  Abajo}}}, \bibinfo{journal}{Nano. Lett.} \textbf{\bibinfo{volume}{11}},
  \bibinfo{pages}{3370} (\bibinfo{year}{2011}).

\bibitem[{\citenamefont{Kinyanjui et~al.}(2012)\citenamefont{Kinyanjui,
  Kramberger, Pichler, Meyer, Wachsmuth, Benner, and
  Kaiser}}]{Kinyanjui/etal:2012}
\bibinfo{author}{\bibfnamefont{M.~K.} \bibnamefont{Kinyanjui}},
  \bibinfo{author}{\bibfnamefont{C.}~\bibnamefont{Kramberger}},
  \bibinfo{author}{\bibfnamefont{T.}~\bibnamefont{Pichler}},
  \bibinfo{author}{\bibfnamefont{J.~C.} \bibnamefont{Meyer}},
  \bibinfo{author}{\bibfnamefont{P.}~\bibnamefont{Wachsmuth}},
  \bibinfo{author}{\bibfnamefont{G.}~\bibnamefont{Benner}}, \bibnamefont{and}
  \bibinfo{author}{\bibfnamefont{U.}~\bibnamefont{Kaiser}},
  \bibinfo{journal}{Europhys. Lett.} \textbf{\bibinfo{volume}{97}},
  \bibinfo{pages}{57005} (\bibinfo{year}{2012}).

\bibitem[{\citenamefont{Grigorenko et~al.}(2012)\citenamefont{Grigorenko,
  Polini, and Novoselov}}]{Grigorenko/etal:2012}
\bibinfo{author}{\bibfnamefont{A.~N.} \bibnamefont{Grigorenko}},
  \bibinfo{author}{\bibfnamefont{M.}~\bibnamefont{Polini}}, \bibnamefont{and}
  \bibinfo{author}{\bibfnamefont{K.~S.} \bibnamefont{Novoselov}},
  \bibinfo{journal}{Nat. Photonics} \textbf{\bibinfo{volume}{6}},
  \bibinfo{pages}{749} (\bibinfo{year}{2012}).

\bibitem[{\citenamefont{Woessner et~al.}(2015)\citenamefont{Woessner,
  Lundeberg, Gao, Principi, Alonso-Gonz\'{a}lez, Carrega, KenjiWatanabe,
  Taniguchi, Vignale, Polini et~al.}}]{Woessner/etal:2015}
\bibinfo{author}{\bibfnamefont{A.}~\bibnamefont{Woessner}},
  \bibinfo{author}{\bibfnamefont{M.~B.} \bibnamefont{Lundeberg}},
  \bibinfo{author}{\bibfnamefont{Y.}~\bibnamefont{Gao}},
  \bibinfo{author}{\bibfnamefont{A.}~\bibnamefont{Principi}},
  \bibinfo{author}{\bibfnamefont{P.}~\bibnamefont{Alonso-Gonz\'{a}lez}},
  \bibinfo{author}{\bibfnamefont{M.}~\bibnamefont{Carrega}},
  \bibinfo{author}{\bibnamefont{KenjiWatanabe}},
  \bibinfo{author}{\bibfnamefont{T.}~\bibnamefont{Taniguchi}},
  \bibinfo{author}{\bibfnamefont{G.}~\bibnamefont{Vignale}},
  \bibinfo{author}{\bibfnamefont{M.}~\bibnamefont{Polini}},
  \bibnamefont{et~al.}, \bibinfo{journal}{Nat. Mater.}
  \textbf{\bibinfo{volume}{14}}, \bibinfo{pages}{421} (\bibinfo{year}{2015}).

\bibitem[{\citenamefont{Liou et~al.}(2015)\citenamefont{Liou, Shie, Chen,
  Breitwieser, Pai, Guo, and Chu}}]{Liou/etal:2015}
\bibinfo{author}{\bibfnamefont{S.~C.} \bibnamefont{Liou}},
  \bibinfo{author}{\bibfnamefont{C.-S.} \bibnamefont{Shie}},
  \bibinfo{author}{\bibfnamefont{C.~H.} \bibnamefont{Chen}},
  \bibinfo{author}{\bibfnamefont{R.}~\bibnamefont{Breitwieser}},
  \bibinfo{author}{\bibfnamefont{W.~W.} \bibnamefont{Pai}},
  \bibinfo{author}{\bibfnamefont{G.~Y.} \bibnamefont{Guo}}, \bibnamefont{and}
  \bibinfo{author}{\bibfnamefont{M.-W.} \bibnamefont{Chu}},
  \bibinfo{journal}{Phys. Rev. B} \textbf{\bibinfo{volume}{91}},
  \bibinfo{pages}{045418} (\bibinfo{year}{2015}).

\bibitem[{\citenamefont{Marinopoulos et~al.}(2004)\citenamefont{Marinopoulos,
  Reining, Rubio, and Olevano}}]{Marinopoulos/etal:2004}
\bibinfo{author}{\bibfnamefont{A.~G.} \bibnamefont{Marinopoulos}},
  \bibinfo{author}{\bibfnamefont{L.}~\bibnamefont{Reining}},
  \bibinfo{author}{\bibfnamefont{A.}~\bibnamefont{Rubio}}, \bibnamefont{and}
  \bibinfo{author}{\bibfnamefont{V.}~\bibnamefont{Olevano}},
  \bibinfo{journal}{Phys. Rev. B} \textbf{\bibinfo{volume}{69}},
  \bibinfo{pages}{245419} (\bibinfo{year}{2004}).

\bibitem[{\citenamefont{Low and Avouris}(2014)}]{Low/Avouris:2014}
\bibinfo{author}{\bibfnamefont{T.}~\bibnamefont{Low}} \bibnamefont{and}
  \bibinfo{author}{\bibfnamefont{P.}~\bibnamefont{Avouris}},
  \bibinfo{journal}{ACS Nano.} \textbf{\bibinfo{volume}{8}},
  \bibinfo{pages}{2086} (\bibinfo{year}{2014}).

\bibitem[{\citenamefont{Gao and Yuan}(2011)}]{Gao/Yuan:2011}
\bibinfo{author}{\bibfnamefont{Y.}~\bibnamefont{Gao}} \bibnamefont{and}
  \bibinfo{author}{\bibfnamefont{Z.}~\bibnamefont{Yuan}},
  \bibinfo{journal}{Solid State Commun.} \textbf{\bibinfo{volume}{151}},
  \bibinfo{pages}{1009} (\bibinfo{year}{2011}).

\bibitem[{\citenamefont{Pisarra et~al.}(2014)\citenamefont{Pisarra, Sindona,
  Riccardi, Silkin, and Pitarke}}]{Pisarra/etal:2014}
\bibinfo{author}{\bibfnamefont{M.}~\bibnamefont{Pisarra}},
  \bibinfo{author}{\bibfnamefont{A.}~\bibnamefont{Sindona}},
  \bibinfo{author}{\bibfnamefont{P.}~\bibnamefont{Riccardi}},
  \bibinfo{author}{\bibfnamefont{V.~M.} \bibnamefont{Silkin}},
  \bibnamefont{and} \bibinfo{author}{\bibfnamefont{J.~M.}
  \bibnamefont{Pitarke}}, \bibinfo{journal}{New J. Phys.}
  \textbf{\bibinfo{volume}{16}}, \bibinfo{pages}{083003}
  (\bibinfo{year}{2014}).

\bibitem[{\citenamefont{Mermin}(1970)}]{Mermin:1970}
\bibinfo{author}{\bibfnamefont{N.~D.} \bibnamefont{Mermin}},
  \bibinfo{journal}{Phys. Rev. B} \textbf{\bibinfo{volume}{1}},
  \bibinfo{pages}{2362} (\bibinfo{year}{1970}).

\bibitem[{\citenamefont{Onida et~al.}(2002)\citenamefont{Onida, Reining, and
  Rubio}}]{Onida/Reining/Rubio:2002}
\bibinfo{author}{\bibfnamefont{G.}~\bibnamefont{Onida}},
  \bibinfo{author}{\bibfnamefont{L.}~\bibnamefont{Reining}}, \bibnamefont{and}
  \bibinfo{author}{\bibfnamefont{A.}~\bibnamefont{Rubio}},
  \bibinfo{journal}{Rev.\ Mod.\ Phys.} \textbf{\bibinfo{volume}{74}},
  \bibinfo{pages}{601} (\bibinfo{year}{2002}).

\bibitem[{\citenamefont{Silkin et~al.}(2004)\citenamefont{Silkin, Chulkov, and
  Echenique}}]{Silkin/Chulkov/Echenique:2004}
\bibinfo{author}{\bibfnamefont{V.~M.} \bibnamefont{Silkin}},
  \bibinfo{author}{\bibfnamefont{E.~V.} \bibnamefont{Chulkov}},
  \bibnamefont{and} \bibinfo{author}{\bibfnamefont{P.~M.}
  \bibnamefont{Echenique}}, \bibinfo{journal}{Phys. Rev. Lett.}
  \textbf{\bibinfo{volume}{93}}, \bibinfo{pages}{176801}
  (\bibinfo{year}{2004}).

\bibitem[{\citenamefont{Yuan and Gao}(2009)}]{Yuan/Gao:2009}
\bibinfo{author}{\bibfnamefont{Z.}~\bibnamefont{Yuan}} \bibnamefont{and}
  \bibinfo{author}{\bibfnamefont{S.}~\bibnamefont{Gao}},
  \bibinfo{journal}{Comp. Phys. Commun.} \textbf{\bibinfo{volume}{180}},
  \bibinfo{pages}{466} (\bibinfo{year}{2009}).

\bibitem[{\citenamefont{Mowbray}(2014)}]{Mowbray:2014}
\bibinfo{author}{\bibfnamefont{D.~J.} \bibnamefont{Mowbray}},
  \bibinfo{journal}{Phys. Status Solidi B} \textbf{\bibinfo{volume}{251}},
  \bibinfo{pages}{2509} (\bibinfo{year}{2014}).

\bibitem[{\citenamefont{Azzolini et~al.}(2017)\citenamefont{Azzolini, Morresi,
  Garberoglio, Calliari, Pugno, Taioli, and Dapor}}]{Azzolini/etal:2017}
\bibinfo{author}{\bibfnamefont{M.}~\bibnamefont{Azzolini}},
  \bibinfo{author}{\bibfnamefont{T.}~\bibnamefont{Morresi}},
  \bibinfo{author}{\bibfnamefont{G.}~\bibnamefont{Garberoglio}},
  \bibinfo{author}{\bibfnamefont{L.}~\bibnamefont{Calliari}},
  \bibinfo{author}{\bibfnamefont{N.~M.} \bibnamefont{Pugno}},
  \bibinfo{author}{\bibfnamefont{S.}~\bibnamefont{Taioli}}, \bibnamefont{and}
  \bibinfo{author}{\bibfnamefont{M.}~\bibnamefont{Dapor}},
  \bibinfo{journal}{Carbon} \textbf{\bibinfo{volume}{118}},
  \bibinfo{pages}{299} (\bibinfo{year}{2017}).

\bibitem[{\citenamefont{Adler}(1962)}]{Adler:1962}
\bibinfo{author}{\bibfnamefont{S.~L.} \bibnamefont{Adler}},
  \bibinfo{journal}{Phys. Rev.} \textbf{\bibinfo{volume}{126}},
  \bibinfo{pages}{413} (\bibinfo{year}{1962}).

\bibitem[{\citenamefont{Wiser}(1963)}]{Wiser:1963}
\bibinfo{author}{\bibfnamefont{N.}~\bibnamefont{Wiser}},
  \bibinfo{journal}{Phys. Rev.} \textbf{\bibinfo{volume}{129}},
  \bibinfo{pages}{62} (\bibinfo{year}{1963}).

\bibitem[{\citenamefont{Kohn and Sham}(1965)}]{Kohn/Sham:1965}
\bibinfo{author}{\bibfnamefont{W.}~\bibnamefont{Kohn}} \bibnamefont{and}
  \bibinfo{author}{\bibfnamefont{L.~J.} \bibnamefont{Sham}},
  \bibinfo{journal}{Phys. Rev.} \textbf{\bibinfo{volume}{140}},
  \bibinfo{pages}{A1133} (\bibinfo{year}{1965}).

\bibitem[{\citenamefont{Rozzi et~al.}(2006)\citenamefont{Rozzi, Varsano,
  Marini, Gross, , and Rubio}}]{Rozzi/etal:2006}
\bibinfo{author}{\bibfnamefont{C.~A.} \bibnamefont{Rozzi}},
  \bibinfo{author}{\bibfnamefont{D.}~\bibnamefont{Varsano}},
  \bibinfo{author}{\bibfnamefont{A.}~\bibnamefont{Marini}},
  \bibinfo{author}{\bibfnamefont{E.~K.~U.} \bibnamefont{Gross}}, ,
  \bibnamefont{and} \bibinfo{author}{\bibfnamefont{A.}~\bibnamefont{Rubio}},
  \bibinfo{journal}{Phys. Rev. B} \textbf{\bibinfo{volume}{73}},
  \bibinfo{pages}{205119} (\bibinfo{year}{2006}).

\bibitem[{\citenamefont{Bergara et~al.}(2003)\citenamefont{Bergara, Silkin,
  Chulkov, and Echenique}}]{Bergara/etal:2003}
\bibinfo{author}{\bibfnamefont{A.}~\bibnamefont{Bergara}},
  \bibinfo{author}{\bibfnamefont{V.~M.} \bibnamefont{Silkin}},
  \bibinfo{author}{\bibfnamefont{E.~V.} \bibnamefont{Chulkov}},
  \bibnamefont{and} \bibinfo{author}{\bibfnamefont{P.~M.}
  \bibnamefont{Echenique}}, \bibinfo{journal}{Phys. Rev. B}
  \textbf{\bibinfo{volume}{67}}, \bibinfo{pages}{245402}
  (\bibinfo{year}{2003}).

\bibitem[{\citenamefont{Pisarra et~al.}(2016)\citenamefont{Pisarra, Sindona,
  Gravina, Silkin, and Pitarke}}]{Pisarra/etal:2016}
\bibinfo{author}{\bibfnamefont{M.}~\bibnamefont{Pisarra}},
  \bibinfo{author}{\bibfnamefont{A.}~\bibnamefont{Sindona}},
  \bibinfo{author}{\bibfnamefont{M.}~\bibnamefont{Gravina}},
  \bibinfo{author}{\bibfnamefont{V.~M.} \bibnamefont{Silkin}},
  \bibnamefont{and} \bibinfo{author}{\bibfnamefont{J.~M.}
  \bibnamefont{Pitarke}}, \bibinfo{journal}{Phys. Rev. B}
  \textbf{\bibinfo{volume}{93}}, \bibinfo{pages}{035440}
  (\bibinfo{year}{2016}).

\bibitem[{\citenamefont{Shishkin and Kresse}(2006)}]{Shishkin/Kresse:2006}
\bibinfo{author}{\bibfnamefont{M.}~\bibnamefont{Shishkin}} \bibnamefont{and}
  \bibinfo{author}{\bibfnamefont{G.}~\bibnamefont{Kresse}},
  \bibinfo{journal}{Phys. Rev. B} \textbf{\bibinfo{volume}{74}},
  \bibinfo{pages}{035101} (\bibinfo{year}{2006}).

\bibitem[{\citenamefont{Chen et~al.}(2010)\citenamefont{Chen, Guo, and
  He}}]{Chen/Guo/He:2010}
\bibinfo{author}{\bibfnamefont{M.}~\bibnamefont{Chen}},
  \bibinfo{author}{\bibfnamefont{G.-C.} \bibnamefont{Guo}}, \bibnamefont{and}
  \bibinfo{author}{\bibfnamefont{L.}~\bibnamefont{He}}, \bibinfo{journal}{J.
  Phys.: Condens. Matter} \textbf{\bibinfo{volume}{22}},
  \bibinfo{pages}{445501} (\bibinfo{year}{2010}).

\bibitem[{\citenamefont{Li et~al.}(2016)\citenamefont{Li, Liu, Chen, Lin, Ren,
  Lin, Yang, and He}}]{Li/Liu/etal:2016}
\bibinfo{author}{\bibfnamefont{P.}~\bibnamefont{Li}},
  \bibinfo{author}{\bibfnamefont{X.}~\bibnamefont{Liu}},
  \bibinfo{author}{\bibfnamefont{M.}~\bibnamefont{Chen}},
  \bibinfo{author}{\bibfnamefont{P.}~\bibnamefont{Lin}},
  \bibinfo{author}{\bibfnamefont{X.}~\bibnamefont{Ren}},
  \bibinfo{author}{\bibfnamefont{L.}~\bibnamefont{Lin}},
  \bibinfo{author}{\bibfnamefont{C.}~\bibnamefont{Yang}}, \bibnamefont{and}
  \bibinfo{author}{\bibfnamefont{L.}~\bibnamefont{He}},
  \bibinfo{journal}{Comput. Mater. Sci.} \textbf{\bibinfo{volume}{112}},
  \bibinfo{pages}{503} (\bibinfo{year}{2016}).

\bibitem[{aba()}]{abacusweb}
\bibinfo{note}{The ABACUS software webpage: \url{http://abacus.ustc.edu.cn}}.

\bibitem[{\citenamefont{Troullier and Martins}(1991)}]{Troullier/Martins:1991}
\bibinfo{author}{\bibfnamefont{N.}~\bibnamefont{Troullier}} \bibnamefont{and}
  \bibinfo{author}{\bibfnamefont{J.~L.} \bibnamefont{Martins}},
  \bibinfo{journal}{Phys. Rev. B} \textbf{\bibinfo{volume}{43}},
  \bibinfo{pages}{1993} (\bibinfo{year}{1991}).

\bibitem[{\citenamefont{Kleinman and Bylander}(1982)}]{Kleinman/Bylander:1982}
\bibinfo{author}{\bibfnamefont{L.}~\bibnamefont{Kleinman}} \bibnamefont{and}
  \bibinfo{author}{\bibfnamefont{D.~M.} \bibnamefont{Bylander}},
  \bibinfo{journal}{Phys. Rev. Lett.} \textbf{\bibinfo{volume}{48}},
  \bibinfo{pages}{1425} (\bibinfo{year}{1982}).

\bibitem[{\citenamefont{Perdew and Zunger}(1981)}]{Perdew/Zunger:1981}
\bibinfo{author}{\bibfnamefont{J.~P.} \bibnamefont{Perdew}} \bibnamefont{and}
  \bibinfo{author}{\bibfnamefont{A.}~\bibnamefont{Zunger}},
  \bibinfo{journal}{Phys.\ Rev.\ B} \textbf{\bibinfo{volume}{23}},
  \bibinfo{pages}{5048} (\bibinfo{year}{1981}).

\bibitem[{\citenamefont{Pines}(1956)}]{Pines:1956}
\bibinfo{author}{\bibfnamefont{D.}~\bibnamefont{Pines}}, \bibinfo{journal}{Can.
  J. Phys.} \textbf{\bibinfo{volume}{34}}, \bibinfo{pages}{1379}
  (\bibinfo{year}{1956}).

\bibitem[{\citenamefont{Principi et~al.}(2014)\citenamefont{Principi, Carrega,
  Lundeberg, Woessner, Koppens, Vignale, and Polini}}]{Principi/etal:2014}
\bibinfo{author}{\bibfnamefont{A.}~\bibnamefont{Principi}},
  \bibinfo{author}{\bibfnamefont{M.}~\bibnamefont{Carrega}},
  \bibinfo{author}{\bibfnamefont{M.~B.} \bibnamefont{Lundeberg}},
  \bibinfo{author}{\bibfnamefont{A.}~\bibnamefont{Woessner}},
  \bibinfo{author}{\bibfnamefont{F.~H.~L.} \bibnamefont{Koppens}},
  \bibinfo{author}{\bibfnamefont{G.}~\bibnamefont{Vignale}}, \bibnamefont{and}
  \bibinfo{author}{\bibfnamefont{M.}~\bibnamefont{Polini}},
  \bibinfo{journal}{Phys. Rev. B} \textbf{\bibinfo{volume}{90}},
  \bibinfo{pages}{165408} (\bibinfo{year}{2014}).

\bibitem[{\citenamefont{Principi et~al.}(2013)\citenamefont{Principi, Vignale,
  Carrega, and Polini}}]{Principi/etal:2013}
\bibinfo{author}{\bibfnamefont{A.}~\bibnamefont{Principi}},
  \bibinfo{author}{\bibfnamefont{G.}~\bibnamefont{Vignale}},
  \bibinfo{author}{\bibfnamefont{M.}~\bibnamefont{Carrega}}, \bibnamefont{and}
  \bibinfo{author}{\bibfnamefont{M.}~\bibnamefont{Polini}},
  \bibinfo{journal}{Phys. Rev. B} \textbf{\bibinfo{volume}{88}},
  \bibinfo{pages}{195405} (\bibinfo{year}{2013}).

\end{thebibliography}
